\documentclass[12pt,preprint,longnamesfirst]{aastex}
\newcommand{\rhobar}{\bar \rho}
\newcommand{\erfc}{{\rm erfc}}
\newcommand{\Label}{\label}

\newcommand{\Citet}{\citet}

\newcommand\plotonefull[3]{%
\centering
\leavevmode
\includegraphics[keepaspectratio, height={#2\columnwidth}, angle=#3]{#1}%
}%

\newcommand\plotoneright[1]{\plotonefull{#1}{1}{90}}

\def\eps@scaling{1.0}%

\newcommand\plottworotated[2]{%
\centering
\leavevmode
\columnwidth=.44\columnwidth
\includegraphics[keepaspectratio=false, width={1.5\columnwidth}, height={\columnwidth}, angle=90]{#1}%
\hfil
\includegraphics[keepaspectratio=false, width={1.5\columnwidth}, height={\columnwidth}, angle=90]{#2}%
}%

\newcommand{\kvec}{{\bf k}}

\newcommand{\rvec}{{\bf r}}

\newcommand{\xvec}{{\bf x}}

\newcommand{\yvec}{{\bf y}}

\newcommand{\xiv}{\hbox to .015em{$\xi$\hss}\hbox to .015em{$\xi$\hss}\xi}
\newcommand{\omvec}{\hbox to .015em{$\omega$\hss}\hbox to .015em{$\omega$\hss}\omega}

\newcommand{\muvec}{\hbox to .015em{$\mu$\hss}\hbox to .015em{$\mu$\hss}\mu}
\newcommand{\betavec}{\hbox to .015em{$\beta$\hss}\hbox to .015em{$\beta$\hss}\beta}

\newcommand{\km}{\, {\rm km}}
\newcommand{\seconds}{\, {\rm sec}}

\newcommand{\Mpc}{\, {\rm Mpc}}
\newcommand{\Hubbleunits}{\, {\rm km}/{\rm sec}/{\rm Mpc}}
\newcommand{\kpc}{\, {\rm kpc}}

\newcommand{\persqarcsec}{\, {\rm arcsec}^{-2}}

\newcommand{\del}{\nabla}
\newcommand{\grad}{\del}

\newcommand{\identicallyeq}{\equiv}
\newcommand{\defeq}{\equiv}

\newcommand{\MSun}{M_{\odot}}

\begin{document}

\title{Gravitational Lensing by Galaxy Groups in the Hubble Deep Field}
\author{David C. Fox}
\affil{Physics Department, Harvard University, Jefferson Physical Laboratories,
Cambridge, MA 02138}
\email{dcfox@cfa.harvard.edu}
\and
\author{Ue-Li Pen}
\affil{Canadian Institute for Theoretical Astrophysics, University of Toronto,
McLennan Labs,
60 St. George Street,
 Toronto, ON, M5S 3H8, Canada}
\email{pen@cita.utoronto.ca}

% \altaffiltext{1}{Also at: Physics Department, Harvard University}
% and Abraham Loeb}
% \affil{Astronomy Department, Harvard University,\\ 
% 60 Garden St., Cambridge MA 02138}
%\altaffiltext{2}{email:aloeb@cfa.harvard.edu}

% \title{Do the Electrons and Ions in X-ray Clusters Share the Same Temperature?}
% \author{David C. Fox\altaffilmark{1}
% \altaffiltext{1}{Also at: Physics Department, Harvard University}
% and Abraham Loeb}
% \affil{Astronomy Department, Harvard University,\\ 
% 60 Garden St., Cambridge MA 02138}
%\altaffiltext{2}{email:aloeb@cfa.harvard.edu}

%NLX% end exclude from vocabulary builder

\begin{abstract}

We investigate strong lensing of galaxies in the Hubble Deep Field by
foreground groups and clusters of galaxies with masses from $10^{ 13 }$
to $10^{ 15 } \, \MSun$.  Over this mass range, groups with the profile
of Navarro, Frenk, \& White (1995) are less efficient than singular
isothermal spheres at producing multiple images of galaxies, by factors
of $5 \times 10^{ - 2 }$ to $10^{ - 3 }$.  This difference highlights
the sensitivity of the lensing cross section to the central density
profile.  Nonetheless, with either profile we find that the expected
number of galaxies lensed by groups in the Hubble Deep Field is at most
$\lesssim 1$, consistent with the lack of clearly identified group lens
systems.

\end{abstract}

% \centerline{to appear in {\it The Astrophysical Journal}, 1997}

\section{INTRODUCTION}

\Citet{NFW95} investigated the structure of clusters of galaxies in
$N$-body simulations of the Standard Cold Dark Matter cosmology.  From
the virial radius, $r_{ \rm vir }$, down to $0.01 r_{ \rm vir }$, they
found that the dark matter profiles of equilibrium clusters followed
the form
\begin{equation}
\rho ( r ) \propto ( r / r_s )^{-1} ( 1 + r / r_s )^{ - 2 }
\label{proportionality}
\end{equation}
characterized by a single scale, $r_s$.  The NFW profile has a central
density cusp, ($\rho \propto r^{-1}$), shallower than the singular
isothermal sphere ($\rho \propto r^{ - 2 }$), but gradually turns over
and becomes steeper ($\propto r^{ - 3 }$) at large radii.  \Citet{DC91}
had earlier found both features in higher resolution simulations of
galaxies, which they fit with a Hernquist (1990) profile,
\begin{equation}
\rho ( r ) \propto ( r / r_s )^{-1} ( 1 + r / r_s )^{ - 3 }.
\end{equation}

\Citet{NFW96} found that the density profiles of equilibrium dark
matter halos spanning a mass range from dwarf galaxies ($3 \times 10^{
11 } \MSun$) to rich clusters ($3 \times 10^{ 15 } \MSun$) could be
described by equation~(\ref{proportionality}).  They suggested that
this profile was a universal form for dark matter halos.  \Citet{NFW97}
found that the NFW profile also fit simulated halos for a variety of
CDM cosmologies.  Other numerical simulations \citep[e.g.,][]{CL96,
BGV98} have generally agreed with the NFW profile.  However, higher
resolution simulations \citep{TBW97, FM97} have suggested somewhat
steeper central slopes, though still shallower than the singular
isothermal sphere.  In the highest resolution simulations
\citep{Moore98, Moore99}, the central density profile is $\propto r^{ -
p }$ with $p = 1.4$--$1.5$.

It is important to test the predicted universal form of dark matter
halos through observations.  There have been a number of tests on
cluster scales.  \Citet{Bartelmann96} showed that gravitational lenses
with the NFW form could produce the radial arcs seen in some clusters,
but noted that the narrowness of observed tangential arcs might pose a
problem.  \Citet{WNB99} found that both the radii and width of
tangential arcs could only be reconciled with the NFW profile if
clusters with low velocity dispersion ($\sim 1000 \km \seconds^{-1}$)
had massive ($\sim 3 \times 10^{ 12 } \, h^{-1} \MSun$) central
galaxies, increasing the slope of the central density profile.
Observations of galaxy number density profiles \citep{Carlberg97}, and
X-ray surface brightness and temperature profiles \citep{MVFS99, NMF99}
seem generally consistent with the NFW profile.  On somewhat smaller
mass scales, \Citet{Mahdavi99} studied a sample of 20 galaxy groups,
and concluded that the NFW profile successfully reproduced their X-ray
emission and the density and velocity distributions of their galaxies. 
On the other hand, the measured rotation curves of dwarf spiral
galaxies imply constant density cores which are inconsistent with the
Hernquist, NFW, and singular isothermal sphere profiles \citep{Moore94,
FloresPrimack94, Burkert95}.  Similar results have been found for low
surface brightness galaxies \citep{KKBP98, Moore99}. Since
observational limits on the gas and stellar masses indicate that these
galaxies are dominated by dark matter, these results pose a serious
challenge to the universality of the NFW profile.  Ejection of baryons
by supernova explosions might lead to formation of a flat core
\citep{NEF96, GS99}, although this suggestion has been criticized as
unrealistic \citep{BurkertSilk97}, especially in the more massive low
surface brightness galaxies \citep{KKBP98, Moore99}.  \citet{KKBP98}
find that profiles with a shallow cusp, $\rho ( r ) \propto r^{ - 0.2
}$, fit both the observations and their simulations.  However, higher
resolution simulations have not confirmed the simulations, instead
showing a steep divergence, $\rho ( r ) \propto r^{ - 1.5 }$
\citep{Moore99}.

Both simulations and observations highlight the need for tests of the
NFW profile over a wide range of masses and radii.  Some tests, such as
strong lensing by clusters, are very sensitive to the mass profile at
small radii, $r \ll r_s$.  Others, such as the X-ray temperature
profiles reported to date, are limited by finite spatial resolution,
and are more useful as probes at intermediate and large radii.

We propose another test of the NFW profile at small radii: statistics
of gravitational lensing by groups.   The dramatic images of lensed
arcs observed in clusters are familiar \citep[e.g.,][]{Kneib96}. 
Groups are less massive than clusters and thus have smaller lensing
cross sections.  However, clusters are rare objects, making up a small
fraction of the matter in the universe, while groups are much more
numerous.  Others have considered lensing of quasars by groups and
clusters \citep{NarayanWhite88, Kochanek95,  FloresPrimack96,
Maoz97,Keeton}.  \Citet{Cooray99} has made predictions of the number of
arcs on the whole sky, but has restricted his attention to clusters
with mass $M \ge 7.5 \times 10^{ 14 } \MSun$, modeled as singular
isothermal spheres.  \Citet{BL00} have estimated the probability of
lensing of galaxies observable with the Next Generation Space Telescope
by singular isothermal spheres of galaxy mass and higher.

\Citet{Keeton}, noting that the NFW profile has a smaller lensing cross
section than the singular isothermal sphere, particularly at lower
masses, has suggested that this lower cross section may explain the
lack of observations of giant arcs in groups as compared to clusters. 
We will see that the difference in cross sections is due to the
difference between the two profiles at small radii. This sensitivity to
the inner part of the density profile makes the statistics of lensing
by groups a potential test of the inner portions of the NFW profile in
groups.  By calculating lensing rates for both the NFW and SIS models,
we hope to span the range of central profiles seen in simulations.

We consider the Hubble Deep Field as a tool for studying lensing by
groups.  The Hubble Deep Field (HDF) provides high-resolution images
and photometry taken by the Hubble Space Telescope in four spectral
bands, F300W, F450W, F606W, and F814W (each labeled according to the
approximate wavelength in nm of the peak spectral sensitivity of its
filter).  The photometry extends to $10 \sigma$ limits of $27$ to $28$
in the $AB$ magnitude system \citep[see, e.g.,][]{Fukugita96}, one to
three magnitudes deeper than current ground-based images
\citep{Williams96}.  This depth allows detection of a large number of
ordinary galaxies at redshifts $> 1$.  Early estimates of the optical
depth for lensing at these redshifts suggested that the Hubble Deep
Field should contain between three and 10 gravitationally lensed
galaxies with multiple images \citep{Hogg96}.

Several groups have identified possible lens systems in the HDF on the
basis of morphologies and colors. \Citet{Hogg96} have applied a
semi-automated technique based on the CLASS survey
\citep[e.g.,][]{Myers95} to identify lens candidates.  They cite a
system with an arc $1.8 \arcsec$ from an elliptical galaxy, and a
possible counter image $1.4 \arcsec$ on the opposite side, as the most
likely of the 24 candidates examined so far, and also mention another
elliptical and arc separated by $0.9 \arcsec$ (but without a counter
image). \Citet{ZMD97} identified three candidates by visual inspection,
including a system of four blue galaxies separated by $\lesssim 1.0
\arcsec$, as well as the two systems mentioned by \citet{Hogg96}. 
However, subsequent spectroscopy with the Low-Resolution Imaging
Spectrograph on the Keck telescope ruled out the first two candidate
systems and provided inconclusive evidence on the third.  This led
\citet{ZMD97} to conclude that there might be no more than one
multiply-imaged galaxy in the HDF down to an $AB$ magnitude of 27 in
F814W, similar to the magnitude of the candidates examined. 
\Citet{CQM99} calculated the number of multiply-imaged galaxies
expected due to lensing by galaxies, modeled as singular isothermal
spheres, and have used the paucity of apparent lensed galaxies to place
a lower limit on the difference, $\Omega_0 - \Lambda$, between the
matter density in the universe, $\Omega_0 \defeq \rho_0 / \rho_c$, and
the cosmological constant, $\Lambda$.

Since foreground galaxies and groups of galaxies can both contribute to
lensing of background galaxies, the actual number of lensed galaxies
would be an upper limit on each contribution. Even allowing for the
possibility of failure to identify some lensed galaxies, this upper
limit appears likely to be quite low.  We want to know if such a limit
can distinguish between NFW and the singular isothermal sphere density
profiles for groups.  We therefore calculate the number of galaxies in
the Hubble Deep Field expected to have multiple images due to lensing
by groups (and clusters) of either profile.  In \S2, we describe the
details of our calculation and our underlying assumptions.  We present
the results in \S3, and discuss how they depend on the choice of lens
profile and cosmology in \S4.  In \S5, we summarize our conclusions and
prospects for future work.

\section{FINDING THE NUMBER OF LENSED GALAXIES}

We discuss the basic elements which determine the lensing probability:
the cosmology, the lens population, and the source galaxy redshifts. 
Finally, we combine these elements to find the number of galaxies in
the Hubble Deep Field expected to have multiple images.

\subsection{Cosmologies}

We consider three different basic cosmologies: a Standard CDM model
with $\Omega_0 = 1$ (SCDM), a low-density open model with $\Omega_0 =
0.3$ (OCDM), and a flat model with $\Omega_0 = 0.3$ and $\Lambda = 0.7$
($\Lambda$CDM).  In all three cases, we assume a Hubble constant, $H_0
= 70 \Hubbleunits$ ($h = 0.7$) and a baryon density of $\Omega_b h^2 =
0.019$ inferred from big bang nucleosynthesis and the deuterium
abundance in quasar absorption line systems \citep{BurlesTytler98}.

\subsection{The Lens Population}

We discuss the abundance of groups and clusters, their density
profiles, and the cross section of an individual lens.   We are
primarily interested in groups, since they are more numerous.  However,
given the universal NFW form of dark matter halos found in simulations,
we treat both groups and clusters as halos in a continuum in mass.

\subsubsection{Mass Function}
\label{mass_function_section}

We calculate the abundance of halos as a function of mass, or mass
function, using the method of \citet{PressSchechter}.  At early times,
the universe is homogeneous with only small density perturbations
$\delta \defeq ( \rho - \bar \rho ) / \bar \rho \ll 1$ with respect to
the mean density, $\bar \rho ( z ) = \rho_0 ( 1 + z )^3$.  The
Press-Schechter method relates these density perturbations at large
$z$, where their evolution can be described by linear theory, to the
highly nonlinear ($\delta \approx 200$) collapsed structures at low
redshifts.  Despite its simplifying assumptions, the Press-Schechter
method reproduces the mass function found in simulations \citep{ECF96}.

The Press-Schechter method is based on the spherical model for collapse
of a homogeneous perturbation, known as the spherical top-hat model. 
Such a perturbation is unaffected by the isotropic universe outside, so
it behaves as a Friedmann universe of its own with mean density
$\rho^\prime$ and expansion rate $H^\prime$. A sufficiently dense
perturbation will be bound, and will turn around after a finite time,
$t_{ \rm ta }$.  Without pressure support, the perturbation would
collapse to infinite density in a collapse time $t_c = 2 t_{ \rm ta }$.
 Due to pressure, the perturbation is instead assumed to reach virial
equilibrium by the same time, $t_c$.

We are interested in the abundance of virialized halos at a given
redshift, $z$.  For a given cosmology, the collapse time of a spherical
perturbation depends on its overdensity, with the densest perturbations
collapsing first.  At early times, when the overdensity of the
perturbation, $\delta^\prime \defeq ( \rho^\prime - \bar \rho ) / \bar
\rho$ is much less than $1$, its evolution can be described by linear
perturbation theory.  While the spherical model describes the nonlinear
evolution of $\delta^\prime$, perturbations can still be characterized
by the extrapolation $\delta^\prime_0$ of $\delta^\prime ( z )$ to $z =
0$ in linear theory.  In the spherical model, perturbations for which
$\delta^\prime_0$ exceeds a threshold $\delta_c ( z )$ will virialize
at redshift $z$.   Thus, $\delta_c ( z )$ relates the overdensity in
linear theory to the highly nonlinear collapse. For the $\Lambda = 0$
case, $\delta_c ( z )$ can be obtained analytically in terms of the age
of universe, $t ( z )$, and $\Omega_0$ (Kochanek 1995).  For the
$\Lambda$CDM case, we calculate its inverse, $z_c ( \delta_c )$,
\citep{ECF96} and solve for $\delta_c$ numerically.

Of course, the universe doesn't consist of a single homogeneous
spherical perturbation.  It is characterized by the overdensity field
$\delta ( \rvec , z )$, where $\rvec$ is the position in a co-moving
coordinate system such that $d \rvec / dt = 0$ corresponds to
unperturbed Hubble flow.  The Press-Schechter model assumes that
$\delta ( \rvec , z )$ is a Gaussian random field at large $z$.  Its
Fourier components, $\delta_\kvec ( z )$ are therefore uncorrelated and
drawn from Gaussian distributions with variance $\left \langle \left |
\delta_\kvec \right |^2 \right \rangle$.

To connect this picture to the spherical collapse model, we smooth
$\delta ( \rvec , z )$ using a spherical top-hat window function of
co-moving radius $R$, or a physical radius of $R / ( 1 + z )$,
corresponding to a mass scale $M = 4 \pi \rho_0 R^3 / 3$.  The
Press-Schechter method assumes that matter at $\rvec$ will become part
of a halo of mass $M$ at redshift $z$, provided that $M$ is the largest
mass scale for which the smoothed overdensity at $\rvec$ exceeds the
threshold from the spherical model.

The mean overdensity, $( \delta M / M ) ( R )$, within a sphere of
co-moving radius $R$ will also have a Gaussian distribution, so the
probability of $\delta M / M$ exceeding $\delta_c$ is readily
calculated.  The variance, $\sigma_R^2$ or $\sigma^2 ( M )$, of the
$\delta M / M$ distribution, is related to the power spectrum, $P ( k )
\defeq \langle \left | \delta_\kvec \right |^2 \rangle$, by
\begin{equation}
\sigma^2 ( M ) = \sigma_R^2 = ( 2 \pi )^{ - 3 } \int_0^\infty 4 \pi k^2
d k \left \langle \left | \delta_\kvec \right |^2 \right \rangle W^2 (
kR )
\label{mass_fluctuations}
\end{equation}
where
\begin{equation}
W ( u ) \defeq 3 \left ( { \sin u - u \cos u \over u^3 } \right ).
\end{equation}

The probability that a given spherical region will collapse depends on
the ratio of $\sigma_R$ to $\delta_c ( z )$.  Thus, we must also
extrapolate $\sigma_R$ to $z = 0$ in linear theory by extrapolating the
power spectrum in equation~(\ref{mass_fluctuations}).  Ignoring factors
independent of $k$ and $z$, the evolution of $P ( k , z )$ is
\begin{equation}
P ( k , z ) \propto P ( k , z_i ) T^2 ( k , z ) D_1^2 ( z )
\end{equation}
where $D_1 ( z )$ is the linear growth function, and $T ( k , z )$,
which describes the scale-dependent evolution, is called the transfer
function.  We assume a scale-invariant primordial power spectrum $P ( k
, z_i ) \propto k$ at high redshift $z_i$.  We use a BBKS
\citep{BBKS86} transfer function with the shape parameter, $\Gamma_{
\rm eff }$, defined by
\begin{equation}
\Gamma_{ \rm eff } = \Omega_0 h \exp \left ( { - \Omega_b ( \Omega_0 +
1 ) \over \Omega_0 } \right ) ,
\end{equation}
of $0.18$ for $\Omega_0 = 0.3$ and $0.65$ for $\Omega_0 = 1$.  We also
consider an $\Omega_0 = 1$ model with a lower value of $0.25$ for
$\Gamma$.  We normalize the power spectrum by fixing $\sigma_R$ on a
scale of $R = 8 \, h^{-1} \Mpc$. For each cosmology, the value of
$\sigma_8$ is chosen to reproduce the observed abundance of clusters as
a function of their X-ray temperature \citep{Pen98}, yielding $\sigma_8
= 0.53$, $0.91$, and $1.0$ for SCDM, OCDM, and $\Lambda$CDM,
respectively.

Finally, the number of halos per co-moving volume, $dn / dM ( M , z )$,
as a function of  the virial mass, $M$, and the redshift, $z$ is
\begin{equation}
{ dn \over dM } ( M , z ) = - { 1 \over \sqrt { 2 \pi } } { \rho_0
\over M } { \delta_c ( z ) \over \sigma ( M ) } { \partial \ln \sigma^2
( M ) \over \partial M } \exp \left ( - { \delta_c^2 ( z ) \over 2
\sigma^2 ( M ) } \right ).
\label{mass_function_equation}
\end{equation}

\begin{figure}
\plottworotated{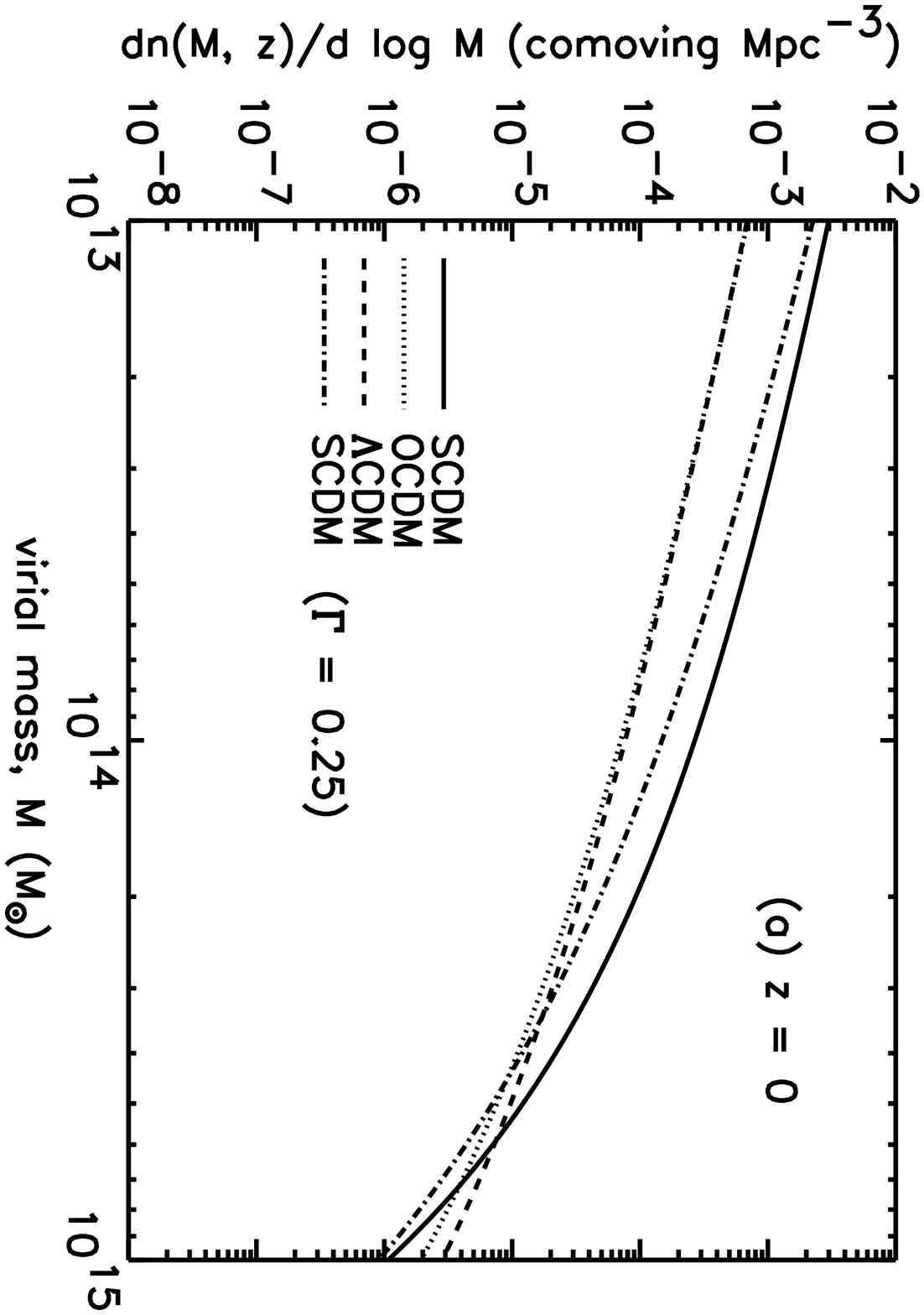}{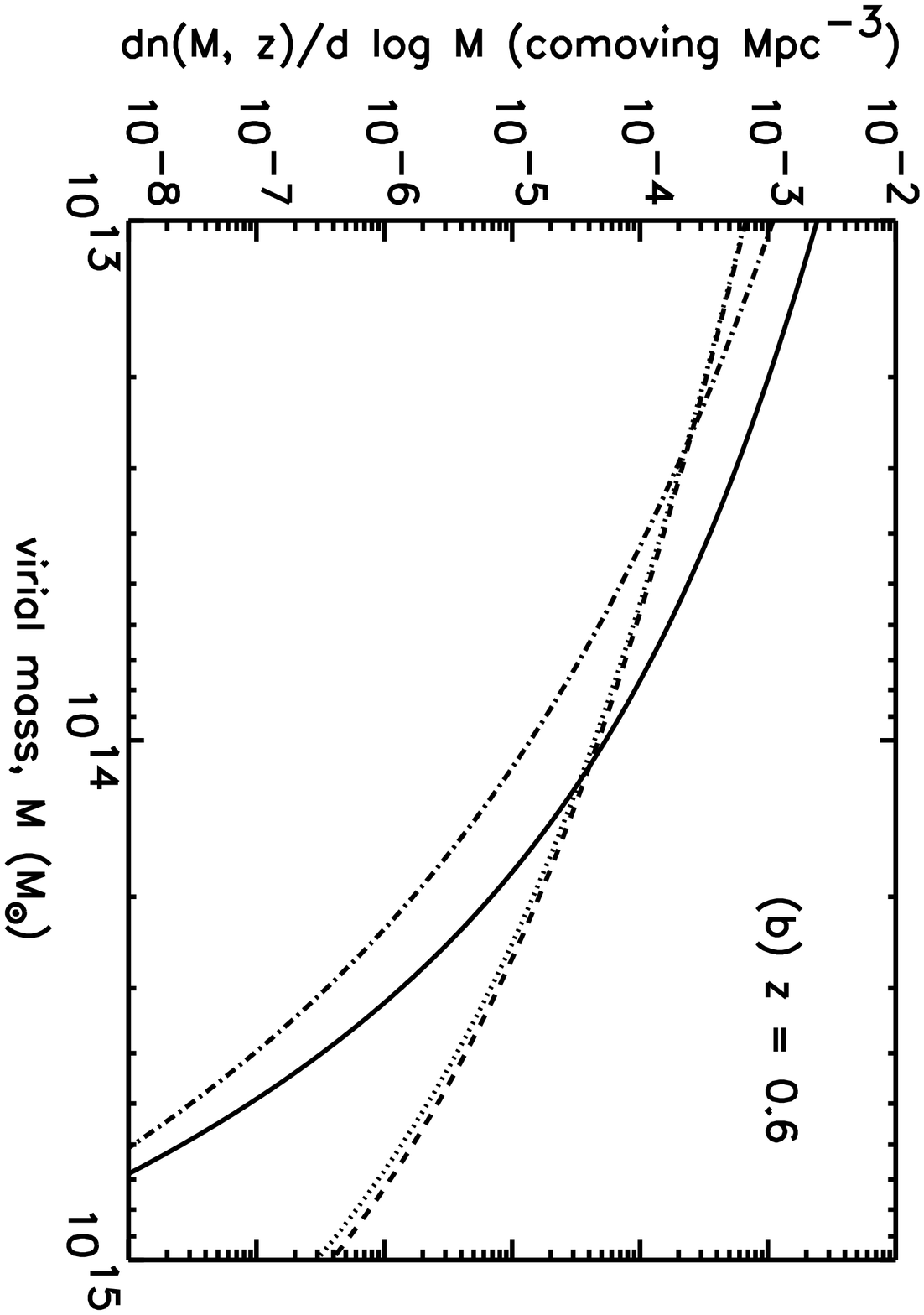}
\caption{Press-Schechter mass function for (a) $z = 0$ and (b) $z =
0.6$ for our three basic cosmologies, Standard CDM (solid line), Open
CDM with $\Omega = 0.3$ (dotted line), and $\Lambda$CDM with $\Omega =
0.3$, $\Lambda = 0.7$ (dashed line), as well as a fourth SCDM variant
with $\Omega = 1$ but $\Gamma = 0.25$ (dot-dashed line).  The power
spectrum has been normalized using the temperature function of clusters
to fix $\sigma_8$ of $0.53$, $0.91$, and $1.0$ for the SCDM (and
variant), OCDM, and $\Lambda$CDM cases,
respectively.\label{mass_function}}
\end{figure}

The resulting mass functions are shown in Figure~\ref{mass_function}
for redshifts of $0$ and $0.6$.  We consider groups and clusters by
restricting our attention to the mass range from $10^{ 13 }$ to $10^{
15 } \, \MSun$. We will see that the number of lensed galaxies
predicted is not sensitive to the upper limit of this range, because of
the exponential cutoff in the mass function for $\sigma^2 ( M ) <
\delta_c^2$.  The lower limit on the mass will have a significant
effect only for the singular isothermal sphere.

\subsubsection{Density Profiles}

We consider two choices for the density profiles of the lens halos, a
singular isothermal sphere (SIS) with
\begin{equation}
\rho ( r ) = { \sigma_v^2 \over 2 \pi G r^2 }
\end{equation}
characterized by a one-dimensional velocity dispersion, $\sigma_v^2$,
and a profile of the form found by \citet{NFW97} with
\begin{equation}
\rho ( r ) = { \rho_s \over ( r / r_s ) ( 1 + r / r_s )^2 }
\end{equation}
characterized by a scale radius, $r_s$, and normalization, $\rho_s =
\delta_{ \rm NFW } \rho_c ( z )$, where $\rho_c ( z )$ is the critical
density at redshift $z$, and we use $\delta_{ \rm NFW }$ instead of
$\delta_c$  to avoid confusion with the overdensity for spherical
collapse.

Our Press-Schechter mass function gives, for each redshift, the
co-moving number density of halos, $d n / dM ( M , z ) \, dM$, with
virial mass between $M$ and $M + dM$.  We need to fix the free
parameters of the density profile in terms of $M$ and $z$.  One
constraint is simply that the profile have the correct virial mass.  In
the spherical top-hat collapse model, the density at virialization,
$\rho_{ \rm vir } \defeq \eta_{ \rm vir } \rhobar ( z )$, can be
calculated using the virial theorem and conservation of energy.  It
depends only on the redshift, $z$, and on the cosmology (see
\citet{ECF96} for the $\Lambda$CDM case).  The virial radius, $r_{ \rm
vir }$, is fixed by $M_{ \rm vir } \defeq 4 \pi \rhobar \, \eta_{ \rm
vir } r_{ \rm vir }^3 / 3$.  Then we can write the virial mass
condition as
\begin{equation}
\left.{ 3 M ( r ) \over 4 \pi r^3 } \right |_{ r = r_{ \rm vir } } =
\rho_{ \rm vir } = \eta_{ \rm vir } \rhobar ( z ).\Label{virial_mass}
\end{equation}

For the singular isothermal sphere, this condition is sufficient to fix
the one free parameter, $\sigma_v^2$.  For the NFW profile, however, we
need additional information.  \Citet{NFW97} have investigated the
relationship between the normalization, $\delta_{ \rm NFW }$, and the
mass $M_{ 200 }$ within a radius $r_{ 200 }$ containing a mean density
of 200 times the critical density $\rho_c ( z )$.  For a wide variety
of cosmologies, they find that $\delta_{ \rm NFW }$ and $M_{ 200 }$ are
well correlated.  While the precise relation depends on cosmology, it
can be fit by 
\begin{equation}
\delta_{ \rm NFW } ( M_{ 200 } , z ) = 3 \times 10^{ 3 } \, \Omega ( z
) \left ( { 1 + z_{ \rm form } \over 1 + z } \right )^3
\end{equation}
where the formation redshift, $z_{ \rm form }$, is defined as the
redshift by which a fraction $f = 10^{ - 2 }$ of the halo mass has
accumulated, according to Press-Schechter formalism of
\citet{LaceyCole93}.  This redshift is given implicitly in terms of the
error function, $\erfc$, by 
\begin{equation}
{ \delta_c ( z_{ \rm form } ) - \delta_c ( z ) \over \sqrt { 2 \Bigl [
\sigma^2 ( fM_{ 200 } ) - \sigma^2 ( M_{ 200 } ) \Bigr ] } } =
\erfc^{-1} \left ( { 1 \over 2 } \right ) ,
\end{equation}
where $\sigma^2 ( M )$ is the variance of the fractional mass
fluctuations on mass scale $M$ in the spherical top hat model, defined
in $\S \ref{mass_function_section}$ . For $f \ll 1$, this simplifies to
\begin{equation}
\delta_c ( z_{ \rm form } ) = \delta_c ( z ) + C^\prime \sigma ( f M_{
200 } ) \label{NFWnormalization}
\end{equation}
where $C^\prime = \sqrt { 2 } \, \erfc^{-1} ( 1 / 2 ) \approx 0.67$.

Note that while \citet{NFW97} referred to $M_{ 200 }$ as the virial
mass and $r_{ 200 }$ as the virial radius, the overdensity with respect
to the critical density $\rho_c ( z_{ \rm vir } )$ at virialization is
only equal to $200$ for Standard CDM.  Unfortunately, the relation
between $M_{ 200 }$ and $M_{ \rm vir }$ depends on the normalization
$\delta_{ \rm NFW }$ of the profile.  We initially approximate $M_{ 200
}$ by $M$ in solving equation~(\ref{NFWnormalization}) for $\delta_{
\rm NFW }$.  Given $\delta_{ \rm NFW }$, we can solve
equation~(\ref{virial_mass}) for the remaining free parameter, $r_s$. 
We then use $\delta_{ \rm NFW }$ and $r_s$ to find a new approximation
for $M_{ 200 }$, and repeat.  This procedure converges rapidly, with
only a few percent change in the parameters on the first iteration, and
tenths of a percent thereafter.

\subsubsection{Properties of the Lenses}

For each density profile, we find the projected surface mass density,
$\Sigma$, as a function of the projected radius from the center of the
lens.  The lens properties are determined by the convergence, $\kappa =
\Sigma / \Sigma_{ \rm cr }$, where
\begin{equation}
\Sigma_{ \rm cr } = { c^2 \over 4 \pi G } { D_s \over D_lD_{ ls } } ,
\end{equation}
and $D_s$, $D_l$, and $D_{ ls }$ are angular diameter distances to the
source, to the lens, and from the lens to the source, respectively. 
For the $\Lambda$CDM case, we calculate the angular diameter distances
according to \citet{CPT92}.

Following \citet{SEF92}, it is convenient to choose an arbitrary length
scale $\xi_0$ in the image plane, and a corresponding scale $\eta_0 = (
D_s / D_l ) \xi_0$ in the source plane.  In units of these length
scales, the source position, $\yvec$, and the image position, $\xvec$,
are related by the lens equation,
\begin{equation}
\yvec = \xvec - \grad \psi ( \xvec ) , \label{vector_lens_equation}
\end{equation}
where the lens potential, $\psi$, satisfies a Poisson equation $\del^2
\psi = 2 \kappa$.  For an axisymmetric lens, $\psi$ depends only on
radius, and the source, lens, and image positions are collinear.  Using
the divergence theorem,
equation~(\ref{vector_lens_equation}) simplifies to the one-dimensional
form
\begin{equation}
y = x - { m ( x ) \over x } , \label{lens_equation}
\end{equation}
where $x$ and $y$ are both signed quantities, and
\begin{equation}
m ( x ) \defeq 2 \int_0^{ | x | } dx^\prime \, x^\prime \kappa (
x^\prime )
\end{equation}
is the dimensionless mass inside a cylinder of dimensionless radius
$\left | x \right |$.

When comparing different profiles, we must use dimensional quantities. 
The dimensional source position, $\eta \defeq \eta_0 y$, and image
positions, $\xi \defeq \xi_0 x$, are related by
\begin{eqnarray}
\eta & = & { D_s \over D_l } \xi \left [ 1 - { M_{ \rm cyl } ( \xi )
\over \pi \Sigma_{ \rm cr } \xi^2 } \right ] \\
& = &
{ D_s \over D_l } \xi \left [ 1 - { \overline \Sigma ( \xi ) \over
\Sigma_{ \rm cr } } \right ].\Label{dimensional_lens_equation}
\end{eqnarray}
where $M_{ \rm cyl } ( \xi )$ is the mass within a cylinder of radius
$\left | \xi \right |$, and $\overline \Sigma ( \xi ) \defeq M_{ \rm
cyl } ( \xi ) / \pi \xi^2$ is the mean surface density within the
cylinder.

\begin{figure}
% \plotonefull{f2.eps}{0.7}{90}
\plotoneright{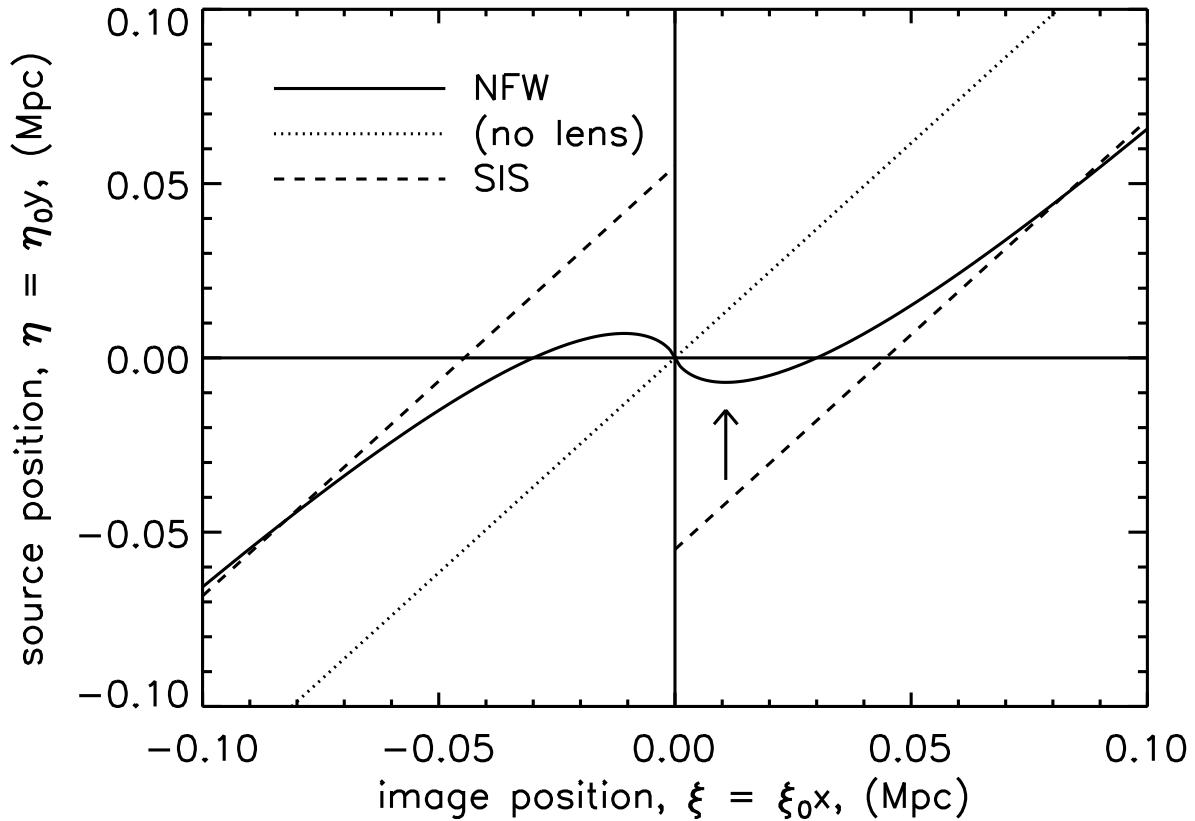}
\caption {Source position, $\eta$, as a function of image position,
$\xi$, for NFW (solid) and SIS (dashed) lenses with $M = 2 \times 10^{
14 } \MSun$, and $z_l = 0.6$, and a source at redshift $z_s = 3$, in an
OCDM cosmology.  The diagonal dotted line shows the relation in the
absence of any lens. A horizontal line of fixed source position,
$\eta$, may intersect each curve at multiple image positions, $\xi$. 
For the NFW lens, the onset of multiple imaging occurs at a local
minimum of $\eta ( \xi )$, indicated by the arrow, producing a radial
caustic crossing.  The SIS profile is sufficiently singular that it has
no local minimum, and the radial caustic is
suppressed.\label{lens_equation_plot}}
\end{figure}

Figure~\ref{lens_equation_plot} shows the lens
equation~(\ref{dimensional_lens_equation}) for NFW and SIS profiles of
the same mass and lens and source redshifts. The particular choice of
mass and redshifts, explained in $\S \ref{comparison_of_profiles}$, is
unimportant here.  A point source at a particular position $\eta_p$
relative to the center of the lens will have an image at each $\xi$
where $\eta ( \xi ) = \eta_p$.  The cross section for multiple imaging
is simply the area of the region of the source plane for which the lens
equation has multiple solutions. 

At locations in the source plane where $d \eta / d \xi = 0$, so that
$\eta ( \xi )$ is tangent to the horizontal, the number of images
changes by two.  Such locations are called radial caustics and the
corresponding curves in the image plane are called radial critical
curves.  If the surface density, $\Sigma$, decreases with radius, the
slope $d \eta / d \xi$ is monotonic increasing with $\xi > 0$. Thus,
there is a single caustic radius, $\eta_c = \eta_0 y_c$, where the
number of images changes from one to three.  Outside this caustic
radius, there is no multiple imaging.  When the source is centered on
the lens, so $\eta = 0$, the two outer images merge to form an Einstein
ring.  The radial critical curve is always inside the Einstein ring
radius, for $\Sigma$ monotonic decreasing.  The singular isothermal
sphere is a special case, because $m ( x ) / x$ is constant and does
not go to $0$ at $x = 0$.  As a result, it has no radial caustic and
its central image is suppressed.

For the singular isothermal sphere, which has no intrinsic length
scale, we make the usual choice of $\xi_0 = 4 \pi ( \sigma_v / c )^2
D_lD_{ ls } / D_s$, set by the ratio of the mass normalization to the
critical density.  Then $\kappa ( x ) = 1 / 2 x$ and $m ( x ) = \left |
x \right |$.  It is easy to show that the lens equation has two image
position solutions when $| y | \le 1$, so the cross section is $\sigma
= \pi \eta_0^2$.

\Citet{Bartelmann96} gives formulas for $\kappa$ and $m$ of the NFW
profile,  for $\xi_0$ chosen equal to the scale radius, $r_s$.  Since
the NFW profile has one length scale, and the ratio $\Sigma_{ \rm cr }
/ \rho_s$ introduces a second, the lens properties no longer scale with
$\xi_0$ the way they do for the singular isothermal sphere.  For each
lens-source pair, we must solve numerically for the caustic radius,
$y_c$, and cross section, $\sigma = \pi y_c^2 \eta_0^2$, for multiple
imaging.

\subsection{The Source Galaxies}
\label{sources}

The lensing cross section depends on the source redshift through the
distances $D_s ( z_s )$ and $D_{ ls } ( z_l , z_s )$ in the critical
density, $\Sigma_{ \rm cr }$.  Several groups \citep{LYF96, SLY97,
WBT98, FLY99} have obtained photometric redshift estimates for the
galaxies in the HDF, with results largely consistent with available
spectroscopic redshifts and with each other \citep{Hogg98}.

We use the \citet{FLY99} catalog of galaxies in the Hubble Deep Field. 
The selection criteria for this catalog are described in detail in
\citet{LYF96} and \citet{FLY99}.  In brief, \citet{FLY99} start by
identifying sources which exceed a minimum surface brightness threshold
in a smooth version of the F814W HDF images.  To include low surface
brightness galaxies, they choose a threshold
$\mu_{ AB ( 8140 ) } = 26.1 \persqarcsec$, corresponding to a
signal-to-noise ratio of only $1.4$ per pixel, but require 10
contiguous pixels above this threshold.  This combination of criteria
is roughly equivalent to an isophotal magnitude limit of $AB ( 8140 ) =
30.6$.  To estimate the completeness limit of the sample, they test the
sensitivity of the identification procedure to the smoothing length and
threshold.  They find that the list of objects identified is
insensitive to these parameters for $A B ( 8140 ) < 28$.  \Citet{FLY99}
therefore restrict their sample to $AB ( 8140 ) < 28.0$.  Near the edge
of the Hubble Deep Field, they impose a stricter magnitude cut of
$26.0$, citing poorer quality images.

\Citet{FLY99} estimate photometric redshifts for the $946$ 
galaxies in the interior and an additional $121$ galaxies near the edge
meeting these criteria.  They supplement the Hubble Space Telescope
images with infrared photometry from the observations of
\citet{Dickinson98} to break degeneracies in the photometric redshift
estimation.   They also include published spectroscopic redshifts for
$108$ of the sources in the catalog.  We use their full catalog of
$1067$ spectroscopic or photometric redshifts as our source redshift
distribution.

\subsection{Putting It All Together}

We discuss the optical depth for lensing, observational selection
effects, magnification bias, and the expected number of multiply-imaged
galaxies in the Hubble Deep Field.

\subsubsection{Optical Depth and Lensing Probability}

The optical depth, $\tau$, for multiple imaging of a source at fixed
redshift, $z_s$, is simply the fraction of the source plane area, $4
\pi D_s^2 ( z_s )$, covered by discs of radius $y_c$ around each lens
position.  The cross section, $\sigma$, of an individual lens is
$\sigma_y \eta_0^2 \defeq \pi y_c^2 \eta_0^2$.  Multiplying by the
number of groups of mass $M$ and redshift $z_l$ in a co-moving volume
$dV_{ \rm com } / dz_l \, dz_l$, we obtain
\begin{eqnarray}
d \tau & = & { \sigma_y \, \eta_0^2 \over 4 \pi D_s^2 } \> { dn \over
dM } ( M , z ) \, dM \, { dV_{ \rm com } \over dz_l } \, dz_l \nonumber
\\
& = & { \sigma_y \, \xi_0^2 \, ( 1 + z_l )^2 \over H ( z_l ) } \> dz_l
\> { dn \over dM } ( M , z ) \, dM.
\label{optical_depth}
\end{eqnarray}
The total optical depth is small, so the lensing probability is equal
to $\tau = \int d \tau$.

\subsubsection{Number of Lensed Galaxies}

We obtain the total number of lensed galaxies expected by multiplying
the number distribution of source galaxies $n_{ \rm gal } ( z_s ) \,
dz_s$ by the lensing probability, $\tau$, and integrating over the
source redshift, $z_s$.  Since we are interested in the number of
lensed galaxies in the Hubble Deep Field, we can use the delta-function
distribution of the discrete catalog of photometric redshifts from
\citet{FLY99} directly, without smoothing, changing
the normalization, or correcting for redshift clustering.

\subsubsection{Observational Selection Effects}

The searches for lensed galaxies in the HDF do not explicitly describe
criteria for identifying lens candidates.  This makes it difficult to
estimate their detection efficiency.  The best we can do is to try to
infer the criteria from the characteristics of the actual candidates
selected.  The lens candidates are selected on the basis of morphology
and color.  Specifically, they include two types of systems, those
consisting of closely spaced galaxies of similar colors, which might be
multiple images of a single source galaxy, and those consisting of a
red elliptical accompanied by a blue arc, which might represent a lens
galaxy and an image of a lens source respectively \citep{ZMD97}.  We
use multiple imaging as a proxy for the latter case as well as the
former.

Several observational selection effects might interfere with the
detection of multiple imaging, thereby reducing the number of lens
systems observed.  The point spread function of the HDF images is
approximately $0.1 \arcsec$ in width, so even closely spaced images
should be easily resolved.  In any case, the typical image separation
increases with lens mass, so spatial resolution is less likely to be a
problem for detecting lensing by groups, even in ground-based images.

A more likely concern is the possibility that lens systems with widely
spaced images might be missed.  Even in the two systems identified by
\citet{ZMD97} as multiple imaging candidates, with separations of $1
\arcsec$ and $3 \arcsec$, spectroscopy showed that the putative
multiple images were actually different galaxies.  The search for lens
candidates may have been limited, explicitly or implicitly, to small
separations, since such false-positive cases would be more likely at
larger separations.

The description of a red elliptical as a possible lens galaxy suggests
that the searches may have focused on lensing by galaxies.  However,
since groups contain galaxies, this might not be a serious bias.

Again, without quantitative criteria, it is difficult to estimate the
reduction in detection efficiency due to such factors.  We therefore
calculate the total number of multiply-imaged galaxies, keeping in mind
that the number actually observed may be lower.

\subsubsection{Magnification Bias}

Another related concern is magnification bias. Lens magnification
increases the angular area of an image, while leaving its surface
brightness unchanged, thereby increasing the flux. If the search for
lensed galaxies is flux limited, sources which would lie below the
detection threshold in the absence of lensing might be detected,
provided the magnification is sufficient.  This magnification bias
differs from other selection effects in that it actually increases the
number of observed lenses.  The extent of the effect depends on the
number of unlensed sources as a function of magnitude, at each
redshift.   For lensing of quasars, where the luminosity function is
steep, magnification bias is a significant effect \citep{Kochanek95}. 
\Citet{Keeton} found that this was particularly important for the NFW
profile, which produces higher mean magnifications than the singular
isothermal sphere.

The key question is whether or not a particular lens survey is flux
limited.  Unlike quasars, galaxies in the Hubble Deep Field are
spatially resolved, so a surface brightness threshold might be a better
detection criterion.  Since lensing does not change surface brightness,
such a survey would have no magnification bias.

On the other hand, some aspects of lens surveys might impose a flux
limit implicitly.  For instance, the smaller the image flux, the harder
it would be to obtain a spectrum of sufficient quality to confirm or
rule out lensing.  \Citet{ZMD97} used the flux of the faintest
component of their three candidates systems to estimate an effective
magnitude limit of $AB ( 8140 ) = 27$.  With only eight components
between these three systems, they do not attempt to prove that their
sample is flux limited.  However, that seems to be their assumption. 
Finally, as discussed in $\S \ref{sources}$, \citet{FLY99} smooth the
HDF images and use an isophotal magnitude limit so as not to exclude
low surface brightness galaxies.  While this does not directly affect
the lensing searches, it is another possible argument in favor of a
flux limit.  Nevertheless, it is difficult to decide whether the lens
searches are better described by a flux limit or by a surface
brightness limit.

To calculate the magnification bias for a flux limited lens survey, we
would need to know the joint distribution with respect to magnitude and
redshift of sources several magnitudes fainter than the effective
magnitude limit of the survey.  However, the photometric redshifts from
the Hubble Deep Field are the deepest redshift survey available,
extending to $AB ( 8140 )$ of $28$.  We could impose a cut on the lens
survey, eliminating candidates with magnitudes too close to $28$, but
this would further reduce the already small number of observed lens
systems.  Alternatively, we could extrapolate the observed
magnitude-redshift distribution to fainter limits.     Since magnitude
and redshift are correlated, the already relatively small redshift
catalog would have to be divided into redshift bands.  Great care would
be required to ensure that the extrapolation was not dominated by noise
due to small number statistics and redshift clustering in the redshift
catalog.  In addition, any such extrapolation beyond the data carries
an unknown risk of systematic error.

Considering these difficulties, we make the simpler choice of assuming
a surface brightness limited sample for lensed galaxies searches in the
HDF.  We therefore neglect the effect of magnification bias on the
expected number of lensed galaxies detected.

Even so, we might be concerned about magnification bias because of the
explicit magnitude cut in the \citet{FLY99} redshift catalog.  However,
the effect on the number of galaxies observed is not the same as the
effect on the number of lensed galaxies observed. The lens
magnification reduces the effective luminosity limit of the
flux-limited sample, but at the same time decreases the surface number
density of sources.  \Citet{BTP95} pointed out that the observed number
of galaxies $N$ per unit solid angle is
\begin{equation}
N_0 \mu^{ \beta ( z ) - 1 }
\end{equation}
where $N_0$ is the number in the absence of lensing, $\mu$ is the local
magnification, and $\beta ( z )$ is minus the logarithmic derivative,
$d \log n ( L , z ) / d \log L$, of the Galaxy luminosity function, $n
( L , z )$, with respect to luminosity.  The slope $\beta$ depends of
course on the source population. It is possible to select source
populations of red galaxies for which $\beta$ differs from unity, but
for the faint blue galaxies which dominate the source population at
high redshifts, $\beta$ is found empirically to be very close to $1$
\citep{Broadhurst95, TDBBK98}.  For such galaxies, the two effects of
magnification cancel, purely by coincidence.  Thus the redshift
distribution is unaffected by lensing.

\section{RESULTS}

In Table~\ref{expected}, we show the total number of lensed galaxies
expected. Groups with singular isothermal sphere profiles produce
roughly $200$ times as many lensed galaxies as those with NFW profiles.
 Even for the SIS lenses, the expected number of lensed galaxies is
$\lesssim 1$.

\begin{figure}
\plotonefull{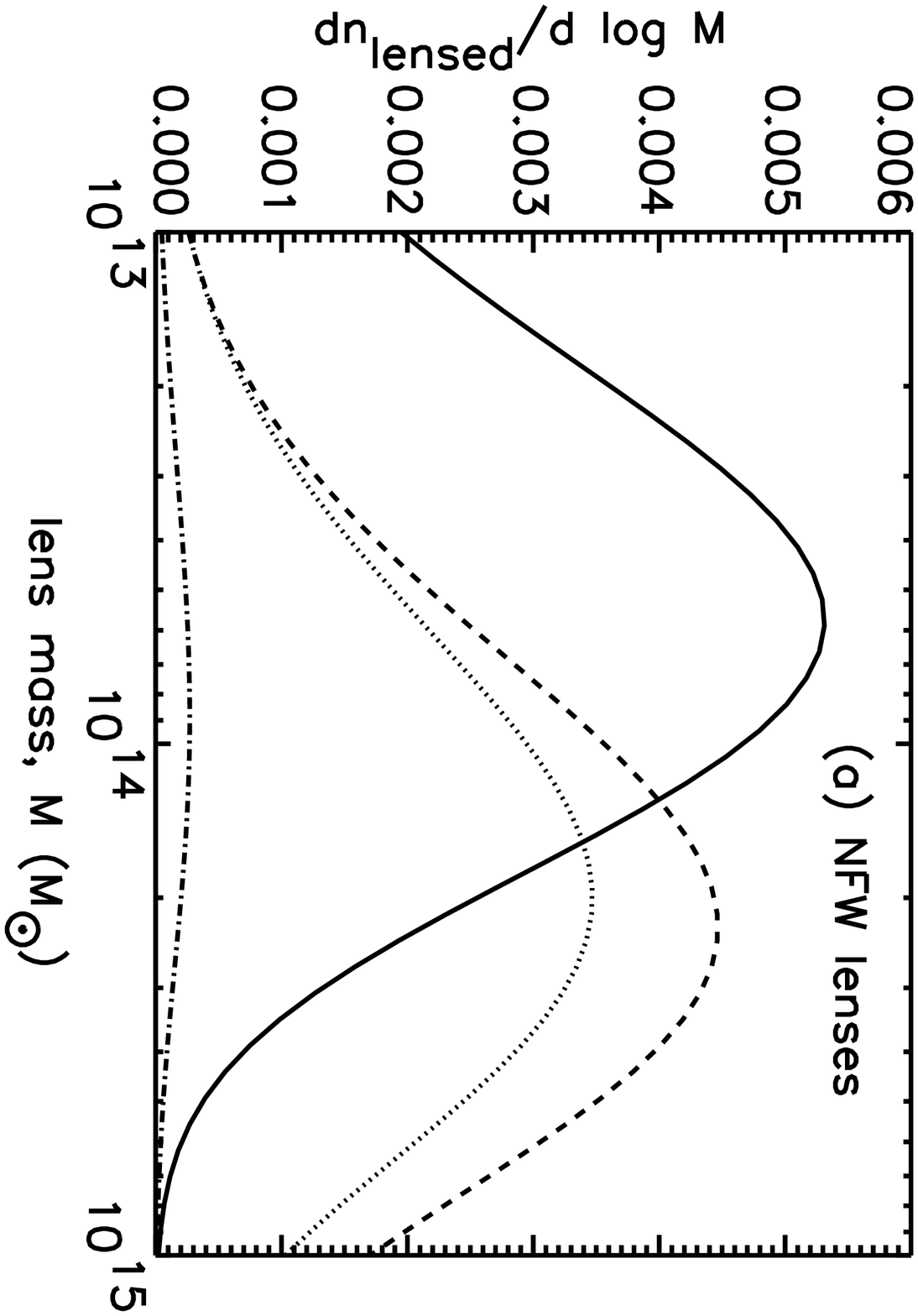}{0.75}{90}
\plotonefull{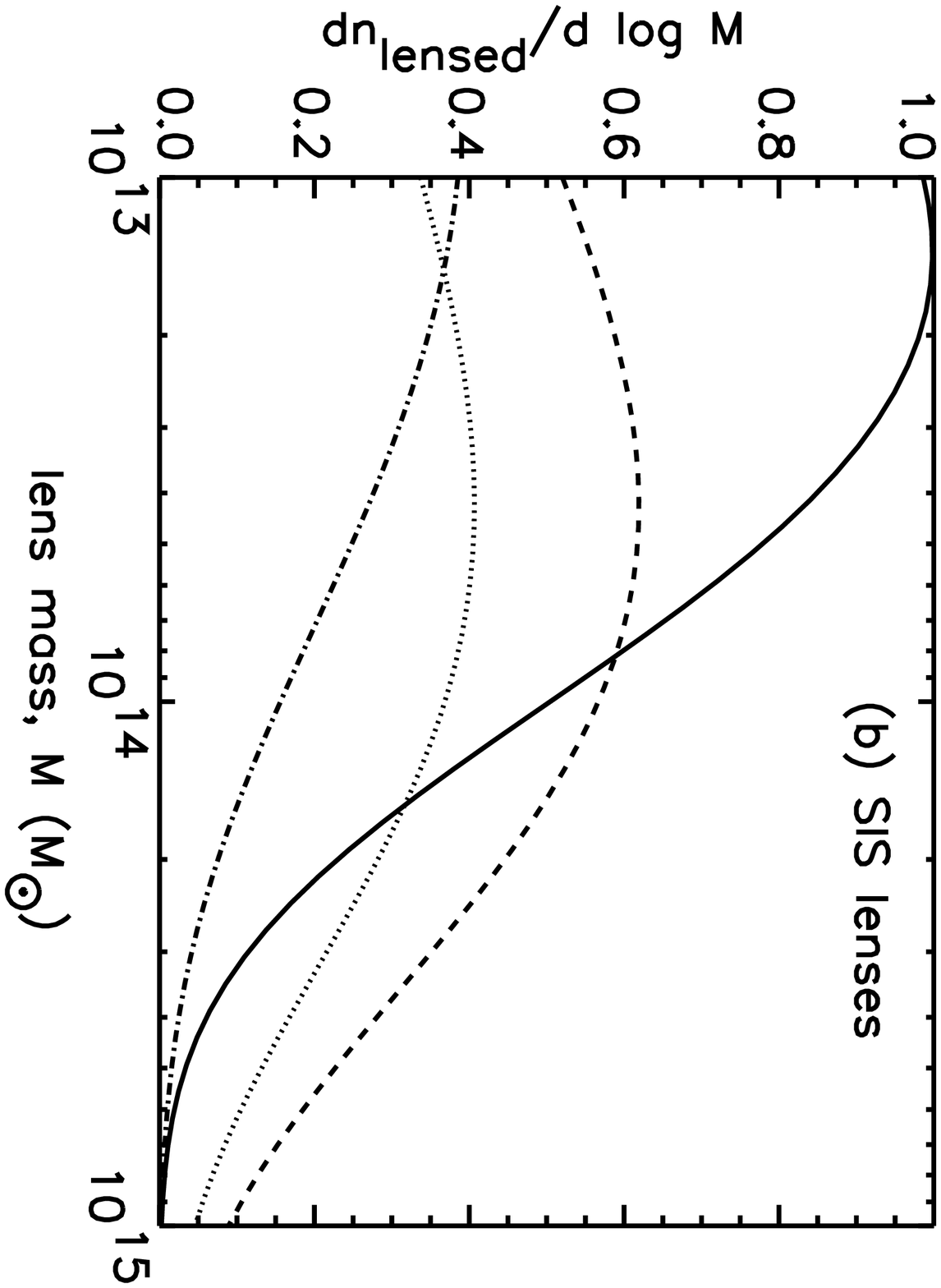}{0.75}{90}
\caption{Number of lensed galaxies expected in the Hubble Deep Field
per unit logarithmic interval in the lens mass, for groups with the (a)
NFW profile and (b) SIS profile.  The cosmologies are the same as in
Figure~\ref{mass_function}.\label{mass_distribution}}
\end{figure}

To understand the results, and to make predictions for future surveys,
we show the distribution of lensed galaxies with respect to various
properties of the lens and source.  In Figure~\ref{mass_distribution}
we show the expected number distribution of lensed galaxies per unit
logarithmic interval in the lens mass.

\begin{figure}
\plotonefull{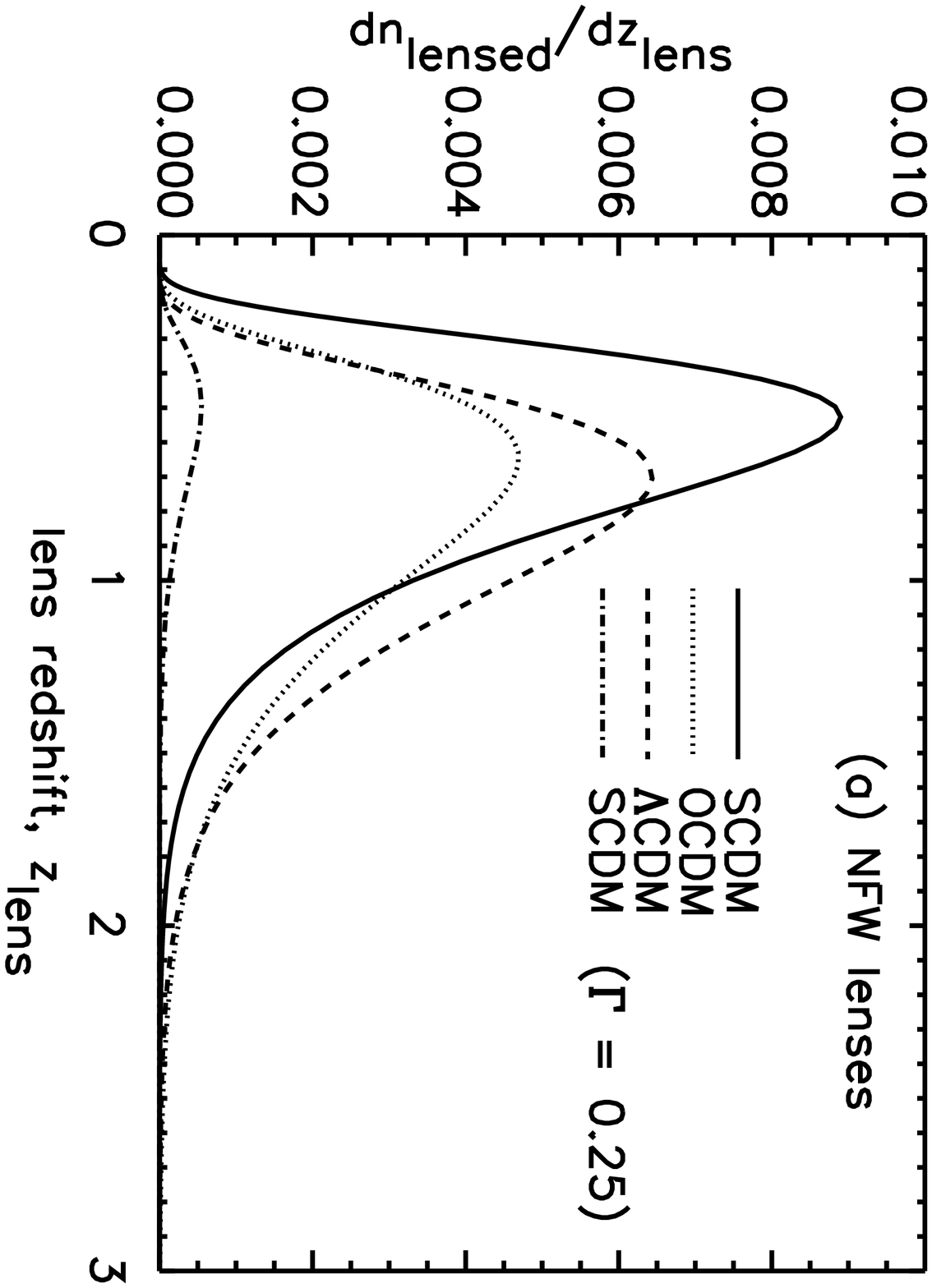}{0.75}{90}
\plotonefull{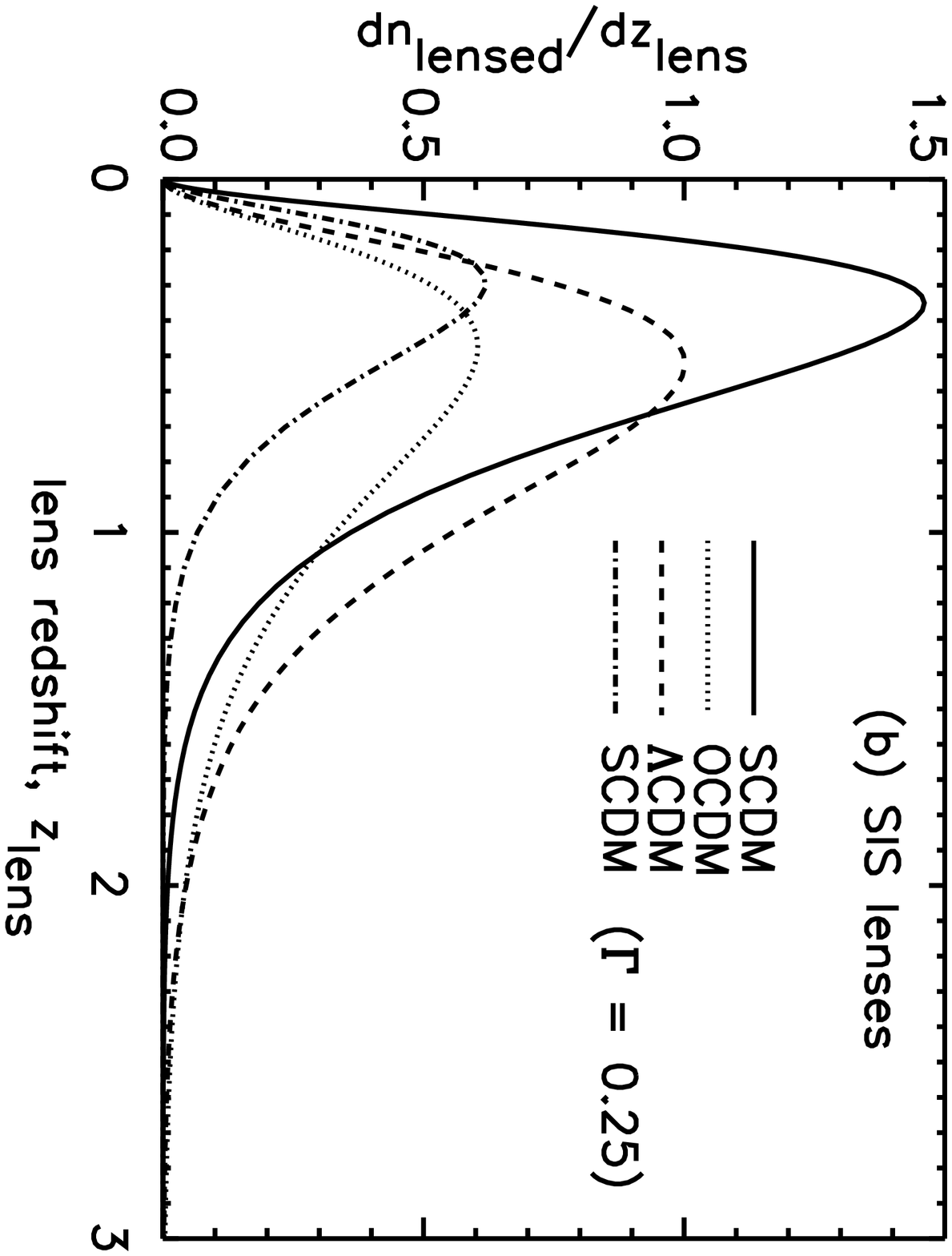}{0.75}{90}
\caption{Number of lensed galaxies in the Hubble Deep Field per unit
interval in lens redshift, for groups with the (a) NFW profile and (b)
SIS profile, again for the same cosmologies as in
Figure~\ref{mass_function}.  \label{lens_redshift}}
\end{figure}

In Figure~\ref{lens_redshift}, we show the distribution of  lenses with
respect to redshift for NFW and SIS groups. The NFW lens distributions
peak near $z_l$ of $0.5$--$0.7$ and are strongly suppressed at $z_l
\lesssim 0.2$.  Compared to the NFW results for the corresponding
cosmology, each SIS lens distribution peaks at a redshift lower by
$\sim 0.2$, and shows only linear suppression at low redshifts.  The
difference at small redshifts reflects the lower lensing efficiency of
the less singular NFW profile when the critical density, $\Sigma_{ \rm
cr }$, is small \citep{Keeton}.  Given sufficient statistics (and
constraints on the cosmology), the difference in lens redshift
distributions is a potential discriminant between NFW and SIS lenses.

\begin{figure}
\plotoneright{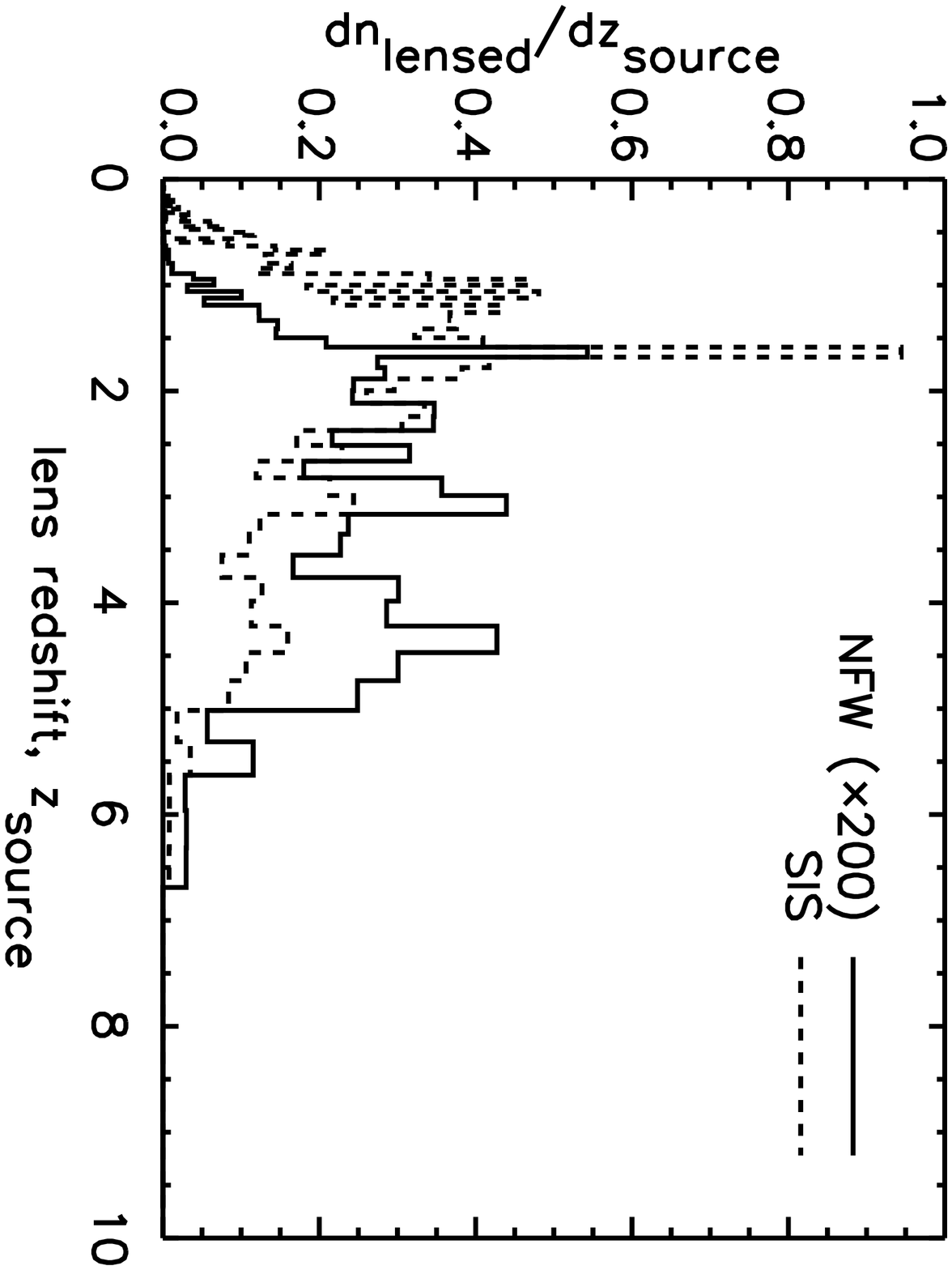}
\caption{Number of lensed galaxies in the Hubble Deep Field per unit
interval in redshift of the lensed galaxies themselves, for the SCDM
case.  Since the total number of lensed galaxies in the SIS case
(dashed) is larger by a factor of $200$, the distribution in the NFW
case (solid) has been scaled up by a factor of $200$ for
comparison.\label{source_redshift}}
\end{figure}

Figure~\ref{source_redshift} shows the redshift distributions of the
lensed galaxies in the SCDM case.  For comparison, the distribution in
the NFW case has been multiplied by $200$, so that it has the same
total area as in the SIS case.  The other three cosmologies have very
similar distributions, apart from the normalization. 
Figure~\ref{source_redshift} reflects the discrete nature of our
catalog-based source redshift distribution.  The distribution of lensed
galaxies in the NFW case is more strongly suppressed at low source
redshift, again reflecting the lower lensing efficiency of the NFW
profile when $\Sigma_{ \rm cr }$ becomes small.

The probability distribution of angular separations between the two
brightest images of lensed galaxies is shown in
Figure~\ref{image_separations}.  Most of the image separations are
greater than $1 \arcsec$, so there should be no difficulty in resolving
the images even with ground-based observations.

\begin{figure}
\plotoneright{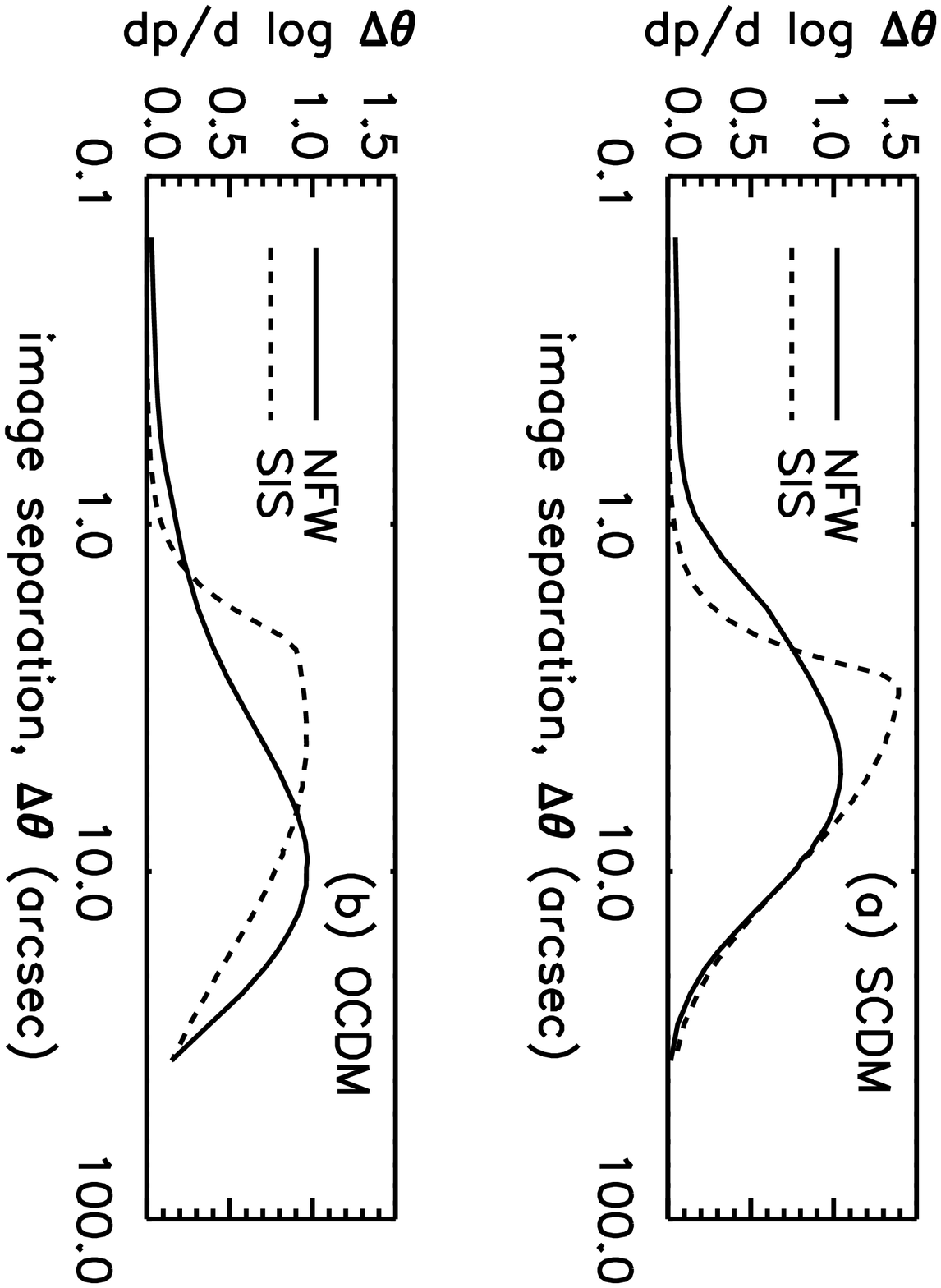}
\caption{Probability distribution of image separations (between the
two brightest images) in arcsec for sources lensed by NFW (solid) and
SIS (dashed) halos, in (a) SCDM ($\Omega = 1$) and (b) OCDM ($\Omega =
0.3$).  The low-$\Gamma$ SCDM and $\Lambda$CDM results are similar to
those for SCDM and OCDM, respectively.\Label{image_separations}}
\end{figure}

\begin{figure}
\plotoneright{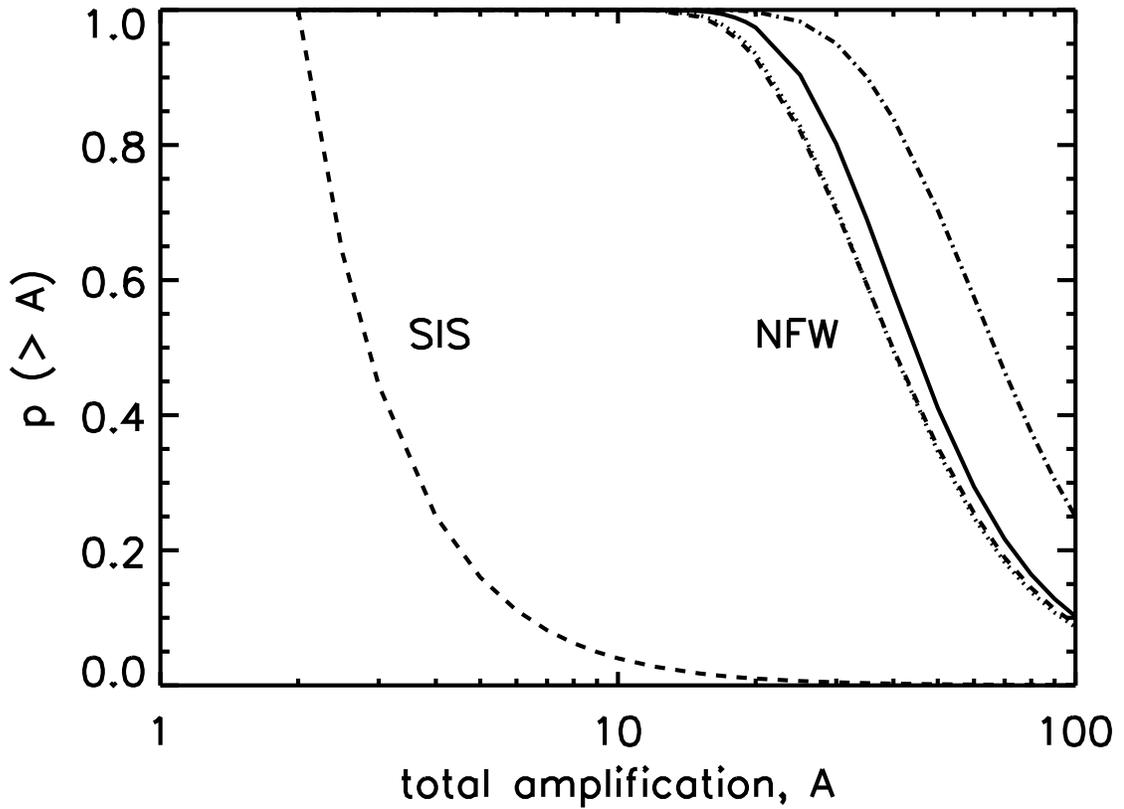}
\caption{Cumulative probability distributions of lensed galaxies
with respect to the minimum total amplification, for NFW lenses (the
four curves at the right, with the same cosmologies as in
Figure~\ref{mass_function}) and SIS lenses (the curve at the left.  The
amplification distribution of the SIS curve, $n ( > A ) = 4 / A^2$, is
independent of cosmology. \label{amplification}}
\end{figure}

Figure~\ref{amplification} shows the cumulative probability
distribution, $p ( > A )$, of lensed galaxies with total amplification
greater than $A$.  The much larger amplifications in the NFW case are
consistent with the results of \citet{Keeton}.

\section{DISCUSSION}

\subsection{NFW vs. SIS}
\label{comparison_of_profiles}

\begin{figure}
\plotoneright{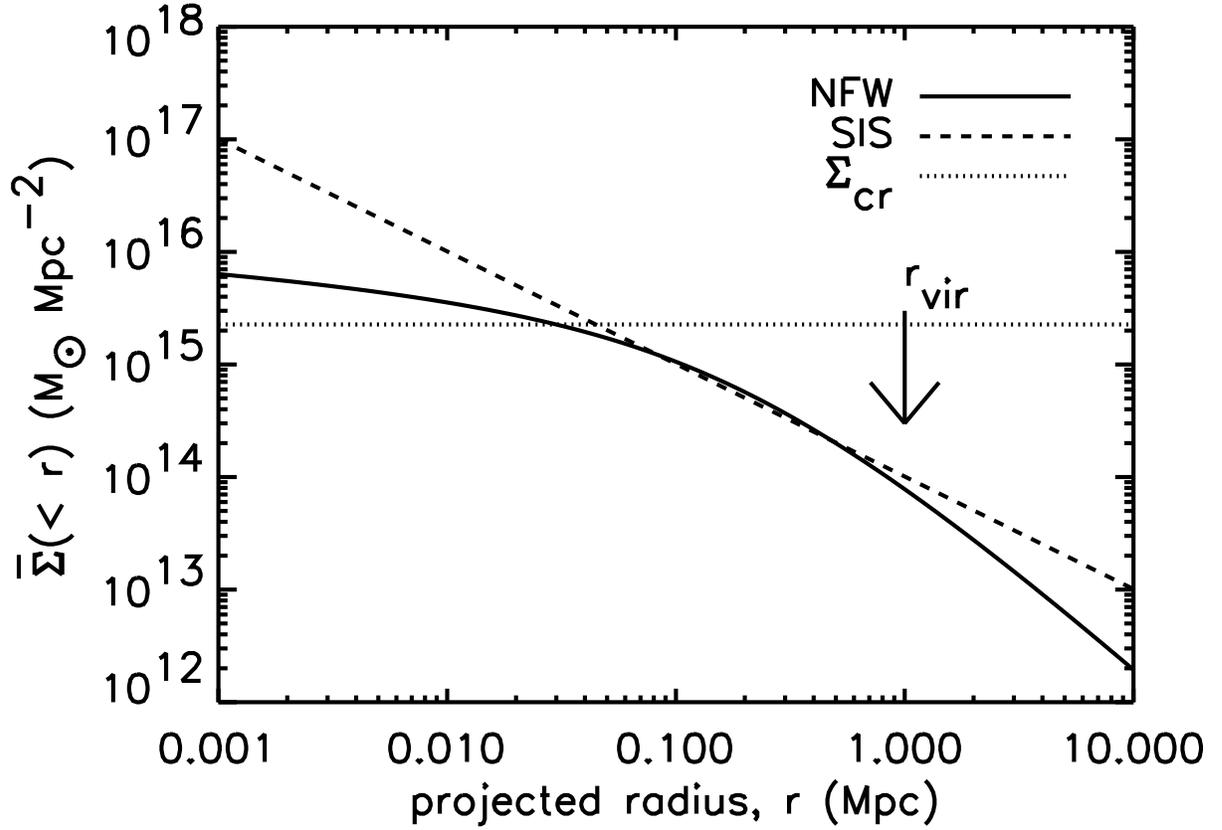}
\caption {Mean surface density, $\overline \Sigma ( r )$ within a
radius $r$, for NFW (solid) and SIS (dashed) profiles of the same mass,
$M = 2 \times 10^{ 14 } \MSun$, and redshift, $z_l = 0.6$, chosen to be
typical of the NFW lenses, in an OCDM cosmology.  The dotted horizontal
line shows the critical surface density, $\Sigma_{ \rm cr }$, assuming
a source redshift $z_s = 3$. It intersects each profile at its Einstein
ring radius. The arrow indicates the location of the virial radius.
\label{surface_density}}
\end{figure}

What makes the NFW profile so much less efficient at lensing than the
singular isothermal sphere?  Both the shape and the normalization of
the profile matter.  To see this, consider a specific example, a halo
with mass $M_{ \rm vir } = 2 \times 10^{ 14 } \MSun$ at $z_l = 0.6$,
typical of our NFW lenses, in an OCDM universe.  Figure
\ref{surface_density} compares the mean surface density, $\overline
\Sigma ( r ) = M_{ \rm cyl } ( r ) / \pi r^2$, within a cylinder of
radius $r$ for NFW and SIS halos of this mass and redshift. Both have
the same mass within the sphere of radius $r_{ \rm vir } \approx 1.0
\Mpc$.  However, the SIS profile has a larger mean surface density both
at small radii ($\lesssim 80 \kpc$) and at large radii ($\gtrsim 450
\kpc$).

At large radii, this is due to the more rapid ($\rho \propto r^{ - 3
}$) falloff of the NFW density profile.  At small radii, the situation
is more complicated.  Our typical NFW lens has a scale radius, $r_s$,
of $0.15 \Mpc$.  For $r \ll r_s$, the NFW profile has a mean surface
density
\begin{equation}
\overline \Sigma_{ \rm NFW } ( r ) \approx 2 \rho_s r_s \ln r_s /
r.\Label{central_surface_density}
\end{equation}
For fixed density, $\rho ( r ) = \rho_s r_s / r$, at small radii,
$\overline \Sigma$ still depends, albeit logarithmically, on the scale
radius.  This occurs because, for density profiles $\rho ( r )$ which
are sufficiently shallow at small radii, $\rho ( r )$ at intermediate
radii $r \lesssim r_s$ can still contribute significantly to $\overline
\Sigma ( r )$ at radii $r \ll r_s$.  For a profile with $\rho \propto
r^{ - p }$ at $r \ll r_s$ and $\rho \propto r^{ - 3 }$ at $r \gg r_s$,
and $1 < p < 2$, the ratio between the contribution to $\overline
\Sigma ( r )$ from intermediate radii and the contribution from small
radii is roughly $( 1 / p ) ( r / r_s )^{ p - 1 }$.  As $p$ approaches
$1$, there will still be a fractional contribution of $O ( \epsilon )$
as long as $r / r_s \gtrsim \epsilon^{ 1 / ( p - 1 ) }$.

Thus, the lower mean surface density of the NFW profile at small radii
is due to the less rapid ($\rho \propto r^{-1}$) divergence of the NFW
profile, but is also influenced by the falloff at large radii.

The key question is which portion of $\overline \Sigma ( r )$
determines the multiple imaging cross section.  The caustic radius,
$\eta_c$, is determined only by the surface mass distribution within
the radial critical curve.  Locating the radial critical curve, $\xi =
\xi_r$, requires solving the lens
equation~(\ref{dimensional_lens_equation}) for $d \eta / d \xi = 0$. 
This occurs where
\begin{equation}
\Sigma_{ \rm cr } = \left.{ d \over d \xi } \left ( \xi \overline
\Sigma \right ) \right |_{ \xi_r } = \left.\overline \Sigma + \xi { d
\overline \Sigma \over d \xi } \right |_{ \xi_r
}.\Label{radial_critical_condition}
\end{equation}
As long as the surface density decreases with radius, $\xi_r$ is always
inside the Einstein ring radius, $r_{ \rm E }$, defined by $\overline
\Sigma ( r_{ \rm E } ) \identicallyeq \Sigma_{ \rm cr }$.  For a
background galaxy at $z_s = 3$, the value of the critical surface
density, $\Sigma_{ \rm cr }$, is shown as a horizontal line in
Figure~\ref{surface_density}, which intersects $\overline \Sigma ( r )$
for the NFW profile at $r = 30 \kpc$.  At all points inside this
radius, $\overline \Sigma$ is smaller for the NFW profile than the
singular isothermal sphere.

To see how this difference in $\overline \Sigma ( r )$ in the lens
plane affects the cross section in the source plane, we need to refer
back to the lens equation~(\ref{dimensional_lens_equation}).  It is
useful to define the offset
\begin{equation}
\Delta ( \xi ) \defeq
\left | \eta - { D_s \over D_l } \xi \right | = { D_s \over D_l }
\left | { \xi \overline \Sigma ( \xi ) \over \Sigma_{ \rm cr } } \right
| \label{offset_definition}
\end{equation}
between the source and image positions (projected into the source
plane).  For the SIS profile, $\overline \Sigma ( \xi ) \propto 1 /
\left | \xi \right |$, so $\Delta ( \xi )$ is independent of $\xi$.

For positive $\xi$, we can solve equation~(\ref{offset_definition}) for
$\xi$ and rewrite the lens equation as
\begin{equation}
\eta = - \Delta ( \xi ) \left ( \Sigma_{ \rm cr } - \overline \Sigma (
\xi ) \right ) / \, \overline \Sigma ( \xi ).
\end{equation}
Evaluating this at the radial critical curve, using
equation~(\ref{radial_critical_condition}), we find
\begin{equation}
| \eta_c | = \left.\Delta ( \xi ) \, { d \log \overline \Sigma \over d
\log \xi } \right |_{ \xi = \xi_r }.
\label{cross_section_offset}
\end{equation}
For the SIS profile, this simplifies to $| \eta_c | = \Delta_{ \rm SIS
}$.  For our typical NFW profile, since $\xi_r < r_{ \rm E }$, and $r_{
\rm E } / r_s \approx 0.2$, we can use
equation~(\ref{central_surface_density}) for $\overline \Sigma$, to
find
\begin{equation}
| \eta_c ( { \rm NFW } ) | = \Delta ( \xi_r ) / \ln ( \xi_0 / \xi_r ) <
\Delta ( \xi_r ).
\end{equation}
Since $\Delta_{ \rm SIS }$ is constant, we can compare it to $\Delta_{
\rm NFW }$ at the critical radius for the NFW lens. The multiple
lensing cross section is $\pi \eta_c^2$, so
\begin{equation}
\sigma_{ \rm NFW } / \sigma_{ \rm SIS } = \left.\left [ { 1 \over \ln (
\xi_0 / \xi ) } { \overline \Sigma_{ \rm NFW } ( \xi ) \over \overline
\Sigma_{ \rm SIS } ( \xi ) } \right ]^2 \> \right |_{ \xi = \xi_r ( {
\rm NFW } ) }.\Label{cross_section_ratio}
\end{equation}
Thus, the smaller cross section for the NFW lens is caused by the
smaller mean surface density, $\overline \Sigma$, as well as by the
smaller proportionality constant between $\eta_c$ and $\Delta ( \xi_r
)$.

The dimensional lens equation~(\ref{dimensional_lens_equation}) is
shown in Figure~\ref{lens_equation_plot} for the particular lens
parameters above.  The range of source positions for which there are
multiple images for the NFW profile is just over one-eighth of that for
the SIS profile, and thus the ratio of lensing cross sections,
$\sigma_{ \rm NFW } / \sigma_{ \rm SIS }$, is $\lesssim 2 \%$, in
agreement with equation~(\ref{cross_section_ratio}).

How do these results depend on mass?  In Figure~\ref{cross_sections},
we show the two cross sections as a function of mass, again for $z_l =
0.6$ and $z_s = 3.0$.  The NFW cross section increases roughly as $M^{
2.25 }$ for this choice of redshifts, whereas the SIS cross section
increases as $M^{ 4 / 3 }$, so $\sigma_{ \rm NFW } / \sigma_{ \rm SIS
}$ increases with mass.  This trend is evident in
Figure~\ref{mass_distribution}.  For the SIS profile, the increase in
$\sigma$ with mass is roughly offset by the decrease in the number of
groups, so the distribution of lens masses is relatively flat or
slightly declining between $10^{ 13 }$ and $10^{ 14 } \MSun$.  Because
the NFW cross section depends more strongly on lens mass than the SIS
cross section, its distribution increases with mass, peaking near $2
\times 10^{ 14 } \MSun$.  The low efficiency of the NFW lenses at small
masses, where halos are most abundant, further reduces the number of
lensed galaxies.  Even for a cluster with $10^{ 15 } \MSun$, $\sigma_{
\rm NFW } / \sigma_{ \rm SIS }$ is still only $\sim 5 \%$, and such
clusters are sufficiently rare in the Press-Schechter model that they
contribute little to the optical depth for lensing.  Our calculated
$\sigma_{ \rm NFW } / \sigma_{ \rm SIS }$ ratios agree with those of
\citet{Keeton}.

\begin{figure}
\plotoneright{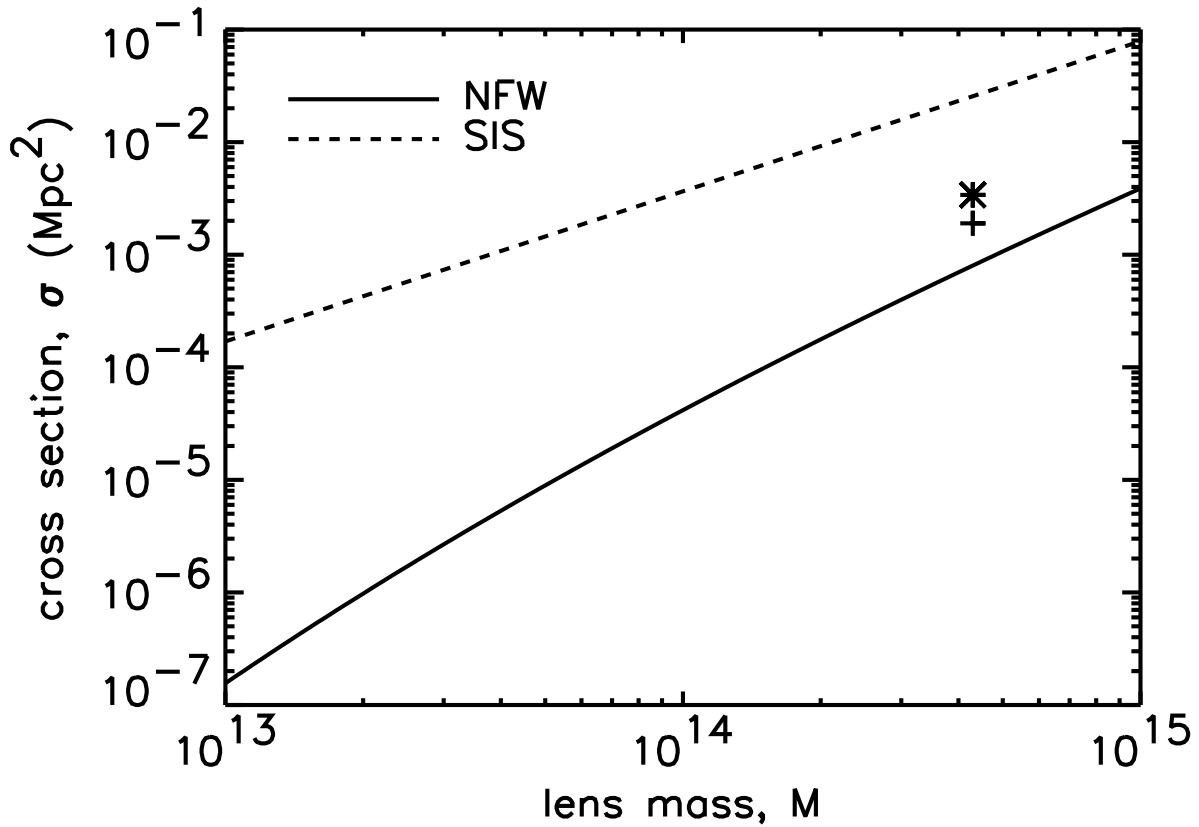}
\caption {Multiple lensing cross sections as a function of mass for
the NFW (solid) and singular isothermal sphere (dashed) profiles, with
lens redshift $z_l = 0.6$ and source redshift $z_s = 3.0$ in an OCDM
cosmology.  The asterisk indicates the cross section for a cluster
described by equation~(\ref{Moore_profile}), from the simulations of
\protect\citet{Moore99}, with $M_{ \rm vir } = 4.3 \times 10^{ 14 } \MSun$ and
$c_{ \rm vir } \defeq r_{ \rm vir } / r_s = 4$. The plus sign indicates
the cross section for a cluster of the same mass, with the
concentration reduced by $20 \%$ in a rough attempt to estimate the
scaling between the $z = 0$ simulation and the assumed lens redshift of
$0.6$ \Label{cross_sections}}
\end{figure}

Given that the highest resolution simulations favor central profiles
$\rho \propto r^{ - p }$ with slope $p \approx 1.5$, is interesting to
ask how the cross section depends on $p$ for $1 < p < 2$. 
Unfortunately, this is not a simple question to answer.  As we have
seen for the NFW lens, the multiple imaging cross section depends not
only on the shape of the mean surface density profile, $\overline
\Sigma ( r )$, but on its normalization at small radii.  For halos
normalized to the same virial mass, fixing $\overline \Sigma ( r )$ at
small radii requires both $p$ and the concentration $c_{ \rm vir }
\defeq r_{ \rm vir } / r_s$, which determines how $\rho ( r )$ turns
over from $r^{ - p }$ at small radii to $r^{ - 3 }$ at large radii. 
For the NFW profile, the relationship of $r_s$ to virial mass has been
investigated extensively through simulations of many halos
\citep{NFW96, NFW97}.  \Citet{Bullock00} have also investigated this
relationship, and have found consistent results at $z = 0$, but lower
mean concentrations at high redshift.  The higher resolution
simulations of \citet{Moore98, Moore99}, who find a density profile of
\begin{equation}
\rho ( r ) \propto { 1 \over \left ( r / r_s \right )^{ 3 / 2 } \left [
1 + \left ( r / r_s \right )^{ 3 / 2 } \right ] } ,
\label{Moore_profile}
\end{equation}
require more computing time, and therefore have been done for only a
few halos, so the dependence of the concentration, $c_{ \rm vir }$, on
the virial mass has not been investigated systematically. 
\Citet{Moore99} do quote a value of $c_{ \rm vir } = 4$ for a single
cluster with virial mass $4.3 \times 10^{ 14 } \MSun$ at $z = 0$ in a
Standard CDM simulation.  The multiple imaging cross section for a halo
of that mass and concentration, but at $z = 0.6$ in our OCDM cosmology,
is shown by the asterisk in Figure~\ref{surface_density}.  This cross
section is $4$ times greater than the corresponding NFW profile, but
still $7.6$ times smaller than the SIS profile.  If, as for an NFW
cluster of the same mass, the concentration declines by $20 \%$ from $z
= 0$ to $z = 0.6$, the cross section would be reduced by $45 \%$, as
shown by the plus sign in Figure~\ref{surface_density}.  Because of the
differences in cosmology and redshift, these comparisons offer at best
a rough idea of the multiple imaging cross section for profiles with
inner slopes intermediate between the NFW profile and a singular
isothermal sphere.  More precise quantitative predictions would require
further high-resolution simulations to determine the variation of the
concentration with halo mass and redshift.

\subsection{Cosmology Dependence}

Consider the distribution of NFW lens masses shown in
Figure~\ref{mass_distribution}.  The distributions for Open CDM
(dotted) and $\Lambda$CDM (dashed) are quite similar up to
normalization.  The $\Lambda$CDM case has about $35 \%$ more lenses,
due to the larger path length and greater angular diameter distance to
a given redshift.  The shape of these distributions reflects the
competition between the increase in $\sigma$ with mass, and the
decreasing abundance of halos.   At low masses, the rapid increase in
the NFW lensing cross section with mass dominates, and the
distributions rise until $\sim 2$--$3 \times 10^{ 14 } \, \MSun$, where
they are truncated  by the high-end cutoff in the mass function (see
Figure~\ref{mass_function}).

Given that Standard CDM has the smallest path length and angular
diameter distances of the three models, we expect it to have the fewest
lenses.   Surprisingly, our calculations indicate slightly more total
lensed galaxies in the SCDM case than in the $\Lambda$CDM case (see
Table~\ref{expected} and Figure~\ref{mass_distribution}).  The reason
can be seen in Figure~\ref{mass_function}.  Compared to the
$\Lambda$CDM and OCDM cases, the SCDM mass function cuts off at a lower
mass due to the lower value of $\sigma_8$. Its lens mass distribution
therefore peaks at a lower mass $\sim 6 \times 10^{ 13 } \MSun$. 
However, the SCDM mass function also has a steeper slope, due to a
larger value of $\Omega_0$ and thus $\Gamma_{ \rm eff }$, resulting in
a {\it larger} number of groups on smaller mass scales (by a factor of
$\approx 3$ for $M = 10^{ 13 } \, \MSun$ and $z = 0.6$) when the power
spectrum is normalized to $\sigma_8$.  To verify this explanation, we
have included a low-$\Gamma$ CDM model with $\Omega = 1$ and $\sigma_8$
still equal to $0.53$, but with $\Gamma = 0.25$.  The low-$\Gamma$
model (dot-dashed), barely visible at the bottom of
Figure~\ref{mass_distribution}, does indeed have far fewer lenses than
the other cosmological models.

Our low mass cutoff at $10^{ 13 } \MSun$ does have an effect on the
total number of lensed galaxies in the SIS case.  Extending the mass
function down to $10^{ 12 } \MSun$ would increase the number of SIS
lenses expected by a factor of $2$ for $\Omega = 1$ and of $1.4$ for
$\Omega = 0.3$.  By contrast, this change would increase the number of
NFW lenses by only 10 percent for SCDM, with a smaller effect for the
other cosmologies.

The difference between our two $\Omega = 1$ models highlights the
sensitivity of predicted lensing by groups to the value of the shape
parameter, $\Gamma$, as previously noted by \citet{Kochanek95}.  With
the power spectrum normalization fixed by the cluster temperature
function, increasing $\Gamma$ produces significantly more groups. 
\Citet{ECF96} note that the normalization of the cluster temperature
function constrains $\sigma_8$ well, since the co-moving radius $R = 8
\, h^{-1} \Mpc$ is close to the cluster scale, but that the slope of
the temperature function, and thus $\Gamma$, is not well determined. 
Measurements of the galaxy correlation function favor $\Gamma \approx
0.25$ \citep{MES96} as does the cluster mass function inferred from the
group luminosity function and an assumed constant mass-to-light ratio
\citep{BahcallCen93}.

As an aside, we note that the \citet{BahcallCen93} mass function has a
similar shape to our Press-Schechter distribution in the OCDM case, but
$40 \%$ more low mass groups at $z = 0$.  However, the Press-Schechter
mass function evolves significantly between $z = 0$ and $z = 0.6$ where
our lens redshift distribution peaks, whereas the \citet{BahcallCen93}
mass function, inferred from low $z$ observations, contains no redshift
evolution information.  As an experiment, we have repeated our
calculation using the \citet{BahcallCen93} mass function and simply
ignoring evolution.  In the OCDM case, with NFW lenses, this yields
$20$ times as many lensed galaxies as the Press-Schechter mass
function, rather than $1.4$ times as many, greatly exaggerating the
number of lensed galaxies expected.  A realistic calculation of lensing
using a group mass function determined observationally at low $z$ would
require the addition of evolution information.

Turning things around, lensing may be able to constrain the group mass
function and thus $\Gamma$. \Citet{Kochanek95} finds that values of
$\Gamma$ between $0.15$ and $0.30$ are also needed to reconcile large
separation quasar lensing statistics with the COBE normalization of
$\sigma_8$ in an SCDM universe.  To constrain $\Gamma$ with lensing of
galaxies by groups, we would need enough lensed galaxies to detect this
effect and disentangle it from the effects of other unknown parameters.
 The small total number of strong lenses expected in the HDF prevents
us from making such inferences.

\section{CONCLUSIONS}

Groups and clusters with the NFW profile are far less efficient at
producing lensed galaxies than singular isothermal spheres of the same
mass.  The ratio, $\sigma_{ \rm NFW } / \sigma_{ \rm SIS }$, of
multiple imaging cross sections ranges from $7 \times 10^{ - 4 }$ for a
$10^{ 13 } \MSun$ group to $5 \times 10^{ - 2 }$ for a $10^{ 15 }
\MSun$ cluster.  The lower efficiency of NFW lenses is a result of
their smaller surface density profile, $\overline \Sigma ( r )$, for
radii much smaller than the scale radius, $r_s$.  This, in turn, is due
to the shallower density profile, $\rho ( r )$, at small radii
(although it also depends logarithmically on $r_s$, which determines
the more rapid falloff in density at large radii).  In principle, this
makes the statistics of multiple imaging a sensitive test of the
central density profile of groups.

Unfortunately, the number of galaxies in the Hubble Deep Field is not
sufficient to perform such a test.  Lensing by SIS groups might be
detected with a sample a few times larger.  However, to detect NFW
groups would require more than $200$ times as many galaxies.  No deep
space-based survey of this size is expected in the near future.
Ground-based surveys would be limited to much lower redshift sources,
and would therefore suffer from smaller optical depths for multiple
imaging.  Still, they might compensate with sufficiently larger fields
containing many more galaxies.  For example the Sloan Digital Sky
Survey (SDSS) is expected to obtain photometry for $5 \times 10^7$
galaxies, as compared to $10^3$ in the Hubble Deep Field.

On the other hand, lensing searches in the HDF illustrate the
difficulties of finding lensed galaxies.  Two of the three candidates
systems identified by morphology and colors were subsequently ruled out
by spectroscopy.  In both cases, hypothetical multiple images turned
out to be different galaxies.  Such a high rate of false positives
limits the number of true lens systems which can be identified in a
given amount of spectroscopic follow-up time. It is unclear whether the
different galaxies were physically associated or not.  In either case,
false positives due to chance superposition are more likely at the
larger image separations expected for lensing by groups.  In contrast,
with lensed quasars, coincidences are negligible, although pairs of
physically associated quasars are still an issue \citep{KFM99}. 
Furthermore, the high luminosity and prominent spectral features of
quasars simplify spectroscopic confirmation.  The SDSS is also expected
to yield a large sample of quasars, which may be a more promising way
of detecting lensing by groups.

Finally, weak lensing statistics may complement strong lensing by
probing the outer portions of the mass profiles of groups.  A number of
detections of weak lensing by galaxies \citep{BBS96, DT96, Hudson98,
Fischer99} and even one early detection of weak lensing by groups
\citep[e.g.,][]{HFK99} have been reported.  We will discuss the
detection and analysis of weak lensing by groups in a subsequent paper.

\acknowledgments

We thank Abraham Loeb for suggesting this project and for his advice on
the work.  We thank William Forman, Irwin Shapiro, and Lars Hernquist
for their advice and support, and for their comments on the paper
drafts, which substantially improved the final result.  We also thank
the anonymous referee, whose suggestions improved the discussion of the
sensitivity of the cross section to the halo profile. This work was
supported by NASA GSRP Fellowship NGT5-50028 for D. C. F and NSERC
Grant 72013704 for U.-L. P.  One of the authors (D. C. F.) would like
to acknowledge Dragon Systems, Inc., whose DragonDictate for Windows
and Dragon NaturallySpeaking software were essential to the development
of the code for this calculation and to the preparation of the paper.

% \clearpage
% \newpage
% for 2-column preprint mode
%\onecolumn

%NLX% exclude from vocabulary builder

\clearpage
\newpage

\begin{table}
\caption {Expected Number of Lensed Galaxies in the HDF}
\label {expected}
\begin {tabular} {l|l|l}
\noalign{\vskip4pt}
lens type & cosmology & number of lenses expected \\
\hline
NFW & SCDM & $5.9 \times 10^{ - 3 }$ \\
& OCDM & $4.1 \times 10^{ - 3 }$ \\
& $\Lambda$CDM & $5.2 \times 10^{ - 3 }$ \\
& SCDM (low $\Gamma$) & $3.1 \times 10^{ - 4 }$ \\
SIS & SCDM & $1.0 \times 10^{0}$ \\
& OCDM & $6.0 \times 10^{-1}$ \\
& $\Lambda$CDM & $9.3 \times 10^{-1}$ \\
& SCDM (low $\Gamma$) & $3.5 \times 10^{-1}$ \\
\end{tabular}
\end{table}

\clearpage
\newpage

%NLX% end exclude from vocabulary builder

%NLX% exclude from vocabulary builder

\begin{thebibliography}{}
\bibitem[Bahcall \& Cen(1993)] {BahcallCen93} Bahcall, N. A., \& Cen,
R.
1993, \apjl, 407, L49
\bibitem[Bardeen et al.(1986)]{BBKS86} Bardeen, J. M., Bond, J. R.,
Kaiser, N., \&
Szalay, A. S. 1986, \apj, 304, 15
\bibitem[Barkana \& Loeb(2000)] {BL00} Barkana, R. \& Loeb, A. 2000,
\apj, 531, 613
\bibitem[Bartelmann(1996)]{Bartelmann96} Bartelmann, M. 1996, \aap,
313,
697
\bibitem[Brainerd et al.(1996)Brainerd, Blandford, \& Smail] {BBS96}
Brainerd, T. G., Blandford, R. D., \& Smail, I. 1996, \apj, 466, 623
\bibitem[Brainerd et al.(1998)Brainerd, Goldberg, Villumsen] {BGV98}
Brainerd,
T. G., Goldberg, D. M., \& Villumsen, J. V. 1998, \apj, 502, 505
\bibitem[Broadhurst(1995)] {Broadhurst95} Broadhurst, T. J. 1995 in AIP
Conf. Proc. 336, Dark Matter, ed.  S. S. Holt \& C. L. Bennett (New
York: AIP)
\bibitem[Broadhurst et al.(1995)Broadhurst, Taylor, \& Peacock] {BTP95}
Broadhurst, T. J., Taylor, A. N., \& Peacock, J. A. 1995, \apj, 438, 49
\bibitem[Bullock et al.(2000)]{Bullock00} Bullock, J. S. et al. 1999,
MNRAS, submitted (astro-ph/9908159)
\bibitem[Burkert(1995)] {Burkert95} Burkert, A. 1995, \apjl, 447, L25
\bibitem[Burkert \& Silk(1997)] {BurkertSilk97} Burkert, A. \& Silk,
J. 1997, \apjl, 488, L55
\bibitem[Burles \& Tytler(1998)] {BurlesTytler98} Burles, S. \& Tytler,
D. 1998, \apj, 507, 732
\bibitem[Carlberg et al.(1997)] {Carlberg97} Carlberg, R. G., et al.
1997, \apjl, 485, L13
\bibitem[Carroll et al.(1992)Carroll, Press, \& Turner] {CPT92}
Carroll, S. M., Press,
W. H., \& Turner, E. L. 1992, \araa, 30, 499
\bibitem[Cole \& Lacey(1996)] {CL96} Cole, S., \& Lacey, C. 1996,
\mnras, 281,
716
\bibitem[Cooray(1999)] {Cooray99} Cooray, A. R. 1999, \aap, 341, 653
\bibitem[Cooray et al.(1999)Cooray, Quashnock, \& Miller] {CQM99}
Cooray, A. R.,
Quashnock, J. M., \& Miller, M. C. 1999, \apj, 511, 562
\bibitem[Dell'Antonio \& Tyson(1996)] {DT96} Dell'Antonio, I. P., \&
Tyson, J. A. 1996, \apjl, 473, L17
\bibitem[Dickinson et al.(1998)] {Dickinson98} Dickinson, M., et al.
1998, in preparation
\bibitem[Dubinski \& Carlberg(1991)] {DC91} Dubinski, J., \& Carlberg,
R. G.
1991, \apj, 378, 496
\bibitem[Eke et al.(1996)Eke, Cole, \& Frenk] {ECF96} Eke, V. R., Cole,
S., \&
Frenk, C. S. 1996, \mnras, 282, 263
\bibitem[Fern\'andez-Soto et al.(1999)Fern\'andez-Soto, Lanzetta, \&
Yahil] {FLY99}
Fern\'andez-Soto, A., Lanzetta, K. M., \& Yahil, A. 1999, \apj, 513, 34
\bibitem[Fischer et al.(1999)] {Fischer99} Fischer, P., et al. 1999,
AJ, submitted (astro-ph/9912119)
\bibitem[Flores \& Primack(1994)] {FloresPrimack94} Flores, R. A., \&
Primack, J. R. 1994, \apjl, 427, L1
\bibitem[Flores \& Primack(1996)] {FloresPrimack96} Flores, R. A., \&
Primack, J. R. 1996, \apjl, 457, L5
\bibitem[Fukugita et al.(1996)] {Fukugita96} Fukugita, M., Ichikawa,
T.,
Gunn, J. E., Doi, M., Shimasaku, K., \& Schneider, D. P. 1996, \aj,
111,
1748
\bibitem[Fukushige \& Makino(1997)] {FM97} Fukushige, T., \& Makino, J.
1997,
\apjl, 477, 9
\bibitem[Gelato \& Sommer-Larsen(1999)] {GS99} Gelato, S., \&
Sommer-Larsen,
J. 1999, \mnras, 303, 321
\bibitem[Hernquist(1990)] {Hernquist90} Hernquist, L. 1990, \apj, 356,
359
\bibitem[Hoekstra et al.(1999)Hoekstra, Franx, \& Kuijken] {HFK99}
Hoekstra, H., Franx, M., \& Kuijken, K. 1999, to appear in
Gravitational Lensing: Recent
Progress and Future Goals ASP conference series, eds. T. Brainerd \& C.
S. Kochanek (astro-ph/9911106) 
\bibitem[Hogg et al.(1996)] {Hogg96} Hogg, D. W., Blandford, R.,
Kundi\'c, T., Fassnacht, C. D., \& Malhotra, S. 1996, \apjl, 467, 73
\bibitem[Hogg et al.(1998)] {Hogg98} Hogg, D. W., et al. 1998, \aj,
115, 1418
\bibitem[Hudson et al.(1998)] {Hudson98} Hudson, M. J., Gwyn, D. J.,
Dahle, H., \& Kaiser, N. 1998, \apj, 503, 531
\bibitem[Keeton(1998)] {Keeton} Keeton, C. R. 1998, doctoral thesis
(Harvard University)
\bibitem[Kneib et al.(1996)] {Kneib96} Kneib, J.-P., Ellis, R. S.,
Smail, I., Couch, W. J., \& Sharples, R. M. 1996, \apj, 471, 643
\bibitem[Kochanek(1995)] {Kochanek95} Kochanek, C. S. 1995, \apj, 453,
545
\bibitem[Kochanek et al.(1999)Kochanek, Falco, \& Mu\~noz] {KFM99}
Kochanek, C. S., Falco, E. E., \& Mu\~noz, J. A. 1999, \apj, 510, 590
\bibitem[Kravtsov et al.(1998)]  {KKBP98} Kravtsov, A. V., Klypin, A.
A.,
Bullock, J. S., \& Primack, J. R. 1998, \apj, 502, 48
\bibitem[Lacey \& Cole(1993)] {LaceyCole93} Lacey, C. \& Cole, S. 1993,
\mnras, 262, 627
\bibitem[Lanzetta et al.(1996)Lanzetta, Yahil, \& Fern\'andez-Soto]
{LYF96} Lanzetta,
K. M., Yahil, A., \& Fern\'andez-Soto, A. 1996, Nature, 381, 759
\bibitem[Mahdavi et al.(1999)] {Mahdavi99} Mahdavi, A., Geller, M.
J., B\"ohringer, H., Kurtz, M. J., \& Ramella, M. 1999, \apj, 518, 69
\bibitem[Maoz et al.(1997)] {Maoz97} Maoz, D., Rix, H.-W., Gal-Yam, A.,
\& Gould, A. 1997, \apj, 486, 75
\bibitem[Markevitch et al.(1999)] {MVFS99} Markevitch, M., Vikhlinin,
A.,
Forman, W. R., \& Sarazin, C. L. 1999,  \apj, 527, 545
\bibitem[Maddox et al.(1996)Maddox, Efstathiou, \& Sutherland] {MES96}
Maddox, S. J.,
Efstathiou, G., \& Sutherland, W. J. 1996, \mnras, 283, 1227
\bibitem[Moore et al.(1994)] {Moore94} Moore, B. 1994, \nat, 370, 629
\bibitem[Moore et al.(1998)] {Moore98} Moore, B., Governato, F., Quinn,
T., 
Stadel, J., \& Lake, G. 1999, \apjl, 499, L5
\bibitem[Moore et al.(1999)] {Moore99} Moore, B., Quinn, T., Governato,
F.,
Stadel, J., \& Lake, G. 1999, \mnras, 310, 1147
\bibitem[Myers et al.(1995)] {Myers95} Myers, S. T., et al. 1995,
\apjl,
447, L5
\bibitem[Narayan \& White(1988)] {NarayanWhite88} Narayan, R. \& White,
S. D. M. 1988, \mnras, 231, P97
\bibitem[Navarro, Eke, \& Frenk(1996)] {NEF96} Navarro, J. F., Eke, V.
R.,
\& Frenk, C. S. 1996, \mnras, 283, L72
\bibitem[Navarro et al.(1995)Navarro, Frenk, \& White]{NFW95} Navarro,
J. F., Frenk,
C. S., \& White, S. D. M. 1995, \mnras, 275, 720
\bibitem[Navarro et al.(1996)Navarro, Frenk, \& White]{NFW96} Navarro,
J. F., Frenk,
C. S., \& White, S. D. M. 1996, \apj, 462, 563
\bibitem[Navarro et al.(1997)Navarro, Frenk, \& White]{NFW97} Navarro,
J. F., Frenk,
C. S., \& White, S. D. M. 1997, \apj, 490, 493
\bibitem[Nevalainen et al.(1999)Nevalainen, Markevitch, \& Forman]
{NMF99} Nevalainen,
J., Markevitch, M., \& Forman, W. 1999, \apj, 526, 1
\bibitem[Pen(1998)] {Pen98} Pen, U. 1998, \apj, 498, 60
\bibitem[Press \& Schechter(1974)] {PressSchechter} Press, W. H. \&
Schechter,
P. 1974, \apjl431, L71
\bibitem[Sawicki et al.(1997)Sawicki, Lin, \& Yee] {SLY97} Sawicki, M.
J., Lin, H., \&
Yee, H. K. C. 1997, \aj, 113, 1
\bibitem[Schneider et al.(1992)Schneider, Ehlers, \& Falco] {SEF92}
Schneider, P.,
Ehlers, J., \& Falco, E. 1992, Gravitational Lenses (Berlin:
Springer-Verlag)
\bibitem[Taylor et al.(1998)] {TDBBK98} Taylor, A. N., Dye, S.,
Broadhurst, T. J., Ben\'itez, N., \& van Kampen, E. 1998, \apj, 501,
539
\bibitem[Tormen et al.(1997)Tormen, Bouchet, \& White] {TBW97} Tormen,
G., Bouchet, F. R., \& White, S. D. M. 1997, \mnras, 286, 865
\bibitem[Wang et al.(1998)Wang, Bahcall, \& Turner] {WBT98} Wang, Y.,
Bahcall, R., \& Turner, E. L. 1998, \aj, 116, 2081
\bibitem[Williams et al.(1999)Williams, Navarro, \& Bartelmann] {WNB99}
Williams, L. L. R., Navarro, J. F., \& Bartelmann, M. 1999, \apj, 527,
535
\bibitem[Williams et al.(1996)] {Williams96} Williams, R. E., et al.
1996, \aj, 112, 1335
\bibitem[Zepf et al.(1997)Zepf, Moustakas, \& Davis] {ZMD97} Zepf, S.
E.,
Moustakas, L. A., \& Davis, M. 1997, \apjl, 474, 1

\end{thebibliography}
\end{document}